\documentclass[A4paper,11pt,1p]{elsarticle}

\makeatletter
\def\ps@pprintTitle{%
 \let\@oddhead\@empty
 \let\@evenhead\@empty
 \def\@oddfoot{\centerline{\thepage}}%
 \let\@evenfoot\@oddfoot}
\makeatother

\usepackage{graphicx}
\usepackage{dcolumn}
\usepackage{bm}
\usepackage{slashed}
\usepackage{hyperref}
\usepackage{amsmath,amssymb}

\newcommand{\mev}{\ensuremath{\mathrm{\,Me\kern -0.1em V}}\xspace}
\newcommand{\mevc}{\ensuremath{{\mathrm{\,Me\kern -0.1em V\!/}c}}\xspace}
\newcommand{\mevcc}{\ensuremath{{\mathrm{\,Me\kern -0.1em V\!/}c^2}}\xspace}
\newcommand{\gev}{\ensuremath{\mathrm{\,Ge\kern -0.1em V}}\xspace}
\newcommand{\gevc}{\ensuremath{{\mathrm{\,Ge\kern -0.1em V\!/}c}}\xspace}
\newcommand{\gevcnospace}{\ensuremath{{\mathrm{\,Ge\kern -0.1em V\!/}c}}}
\newcommand{\gevcc}{\ensuremath{{\mathrm{\,Ge\kern -0.1em V\!/}c^2}}\xspace}
\newcommand{\bea}{\begin{eqnarray}}
\newcommand{\eea}{\end{eqnarray}}

\def\beq{\begin{equation}}
\def\eeq{\end{equation}}
\def\bea{\begin{eqnarray}}
\def\eea{\end{eqnarray}}

\usepackage{color}

\def\missET {\slashed{E}_T}
\def \bm#1{\mbox{\boldmath$#1$\unboldmath}} 
\def \myeq#1{(\ref{#1})}

\begin{document} 

\begin{frontmatter} 


\title{Simplified Models for Dark Matter Searches at the LHC}

\author[AS]{Jalal Abdallah}
\author[IMP]{Henrique Araujo}
\author[LY1,LYN,CERN]{Alexandre Arbey}
\author[TEL]{Adi Ashkenazi} 
\author[SOU]{Alexander Belyaev}
\author[SLAC]{Joshua Berger}
\author[DUR]{Celine Boehm} 
\author[CERN]{Antonio Boveia}
\author[MEL]{Amelia Brennan}
\author[BRI]{Jim Brooke}
\author[IMP]{Oliver Buchmueller}
\author[RUT]{Matthew Buckley\corref{thanks}}
\author[SIS]{Giorgio Busoni\corref{thanks}}
\author[CAS,BRU1]{Lorenzo Calibbi\corref{thanks}}
\author[DAV]{Sushil Chauhan}
\author[BRU2]{Nadir Daci}
\author[IMP]{Gavin Davies} 
\author[BRU2]{Isabelle De Bruyn} 
\author[NIKHEF]{Paul De Jong}
\author[CERN]{Albert De Roeck} 
\author[IMP]{Kees de Vries} 
\author[ROME]{Daniele Del Re}
\author[SIS]{Andrea De Simone} 
\author[FREI]{Andrea Di Simone}
\author[GEN]{Caterina Doglioni}
\author[SLAC]{Matthew Dolan}
\author[BONN]{Herbi K. Dreiner}
\author[CERN,KCL]{John Ellis} 
\author[MAR]{Sarah Eno}
\author[TEL]{Erez Etzion}
\author[KCL]{Malcolm Fairbairn}
\author[OXF]{Brian Feldstein}
\author[BRI]{Henning Flaecher} 
\author[ARG]{Eric Feng} 
\author[FNAL]{Patrick Fox}
\author[LPSC]{Marie-H\'el\`ene Genest}
\author[UCSB]{Loukas Gouskos} 
\author[GEN]{Johanna Gramling} 
\author[CERN,OXF]{Ulrich Haisch\corref{thanks}}
\author[FNAL]{Roni Harnik}
\author[OXF]{Anthony Hibbs}
\author[MAL]{Siewyan Hoh}
\author[ORE]{Walter Hopkins}
\author[HAR]{Valerio Ippolito}
\author[GEN]{Thomas Jacques\corref{thanks}}
\author[DESY]{Felix Kahlhoefer\corref{thanks}}
\author[DUR]{Valentin V. Khoze}
\author[RHUL]{Russell Kirk\corref{thanks}}
\author[UCL]{Andreas Korn}
\author[OSU]{Khristian Kotov}
\author[TTU]{Shuichi Kunori}
\author[BRO]{Greg Landsberg} 
\author[GRAP]{Sebastian Liem}
\author[CHI,KAV]{Tongyan Lin\corref{thanks}}
\author[BRU2]{Steven Lowette}
\author[IMP,RAL]{Robyn Lucas}
\author[CERN]{Luca Malgeri}
\author[IMP]{Sarah Malik}
\author[DUR,GRAP]{Christopher McCabe} 
\author[UCI]{Alaettin Serhan Mete}
\author[GEN]{Enrico Morgante\corref{thanks}}
\author[FNAL]{Stephen Mrenna}
\author[CERN,KEK]{Yu Nakahama}
\author[BRI]{Dave Newbold} 
\author[GLA]{Karl Nordstrom}
\author[NIKHEF]{Priscilla Pani}
\author[BERK,LBNL]{Michele Papucci} 
\author[WUP]{Sophio Pataraia} 
\author[CHI]{Bjoern Penning} 
\author[ZUR]{Deborah Pinna}
\author[PAV]{Giacomo Polesello}
\author[GEN]{Davide Racco}
\author[OXF]{Emanuele Re}
\author[GEN]{Antonio Walter Riotto}
\author[SLAC]{Thomas Rizzo}
\author[NIKHEF,GRAP]{David Salek}
\author[OXF]{Subir Sarkar}
\author[TOR]{Steven Schramm}
\author[OK]{Patrick Skubic}
\author[TEL]{Oren Slone} 
\author[HEI]{Juri Smirnov\corref{thanks}}
\author[REH]{Yotam Soreq} 
\author[IMP]{Timothy Sumner} 
\author[UCI]{Tim M.~P. Tait\corref{thanks}}
\author[SOU,RAL]{Marc Thomas} 
\author[RAL]{Ian Tomalin}
\author[NIKHEF]{Christopher Tunnell}
\author[CERN]{Alessandro Vichi} 
\author[TEL]{Tomer Volansky}
\author[NYU]{Neal Weiner}
\author[RHUL]{Stephen M. West}
\author[RAL]{Monika Wielers}
\author[RAL]{Steven Worm\corref{thanks}\corref{corr}}
\author[PI,MCM]{Itay Yavin} 
\author[BRU1]{Bryan Zaldivar}
\author[UCI]{Ning Zhou} 
\author[BERK,LBNL]{Kathryn Zurek}
%
\cortext[thanks]{Primary contributor}
\cortext[corr]{Corresponding author}
%
\address[AS]{Academia Sinica Institute of Physics, Taipei 11529, Taiwan}
\address[IMP]{Imperial College London High Energy Physics, London SW7 2AZ, United Kingdom}
\address[LY1]{Universit\'e Lyon 1, Centre de Recherche Astrophysique de Lyon, 69561 Saint-Genis Laval, France}
\address[LYN]{Ecole Normale Sup\'erieure de Lyon, Lyon, France}
\address[CERN]{Physics Department, CERN, Geneva CH-1211 Switzerland}
\address[TEL]{Tel Aviv University Department of Physics, P.O. Box 39040, Tel Aviv 6997801, Israel}
\address[SOU]{University of Southampton Physics and Astronomy, Southampton SO17 1BJ, United Kingdom}
\address[SLAC]{SLAC National Accelerator Laboratory, Menlo Park 94025, USA}
\address[DUR]{Institute for Particle Physics Phenomenology, Durham University, Durham DH1 3LE, United Kingdom}
\address[MEL]{University of Melbourne, Victoria 3010, Australia}
\address[BRI]{HH Wills Physics Laboratory, Tyndall Avenue, Bristol BS8 1TH, United Kingdom}
\address[RUT]{Rutgers University Department of Physics and Astronomy, Piscataway, 08854-8019, USA}
\address[SIS]{SISSA and INFN, Sezione di Trieste, Trieste 34136, Italy}
\address[CAS]{Institute of Theoretical Physics, Chinese Academy of Sciences, Beijing 100190, P.~R.~China}
\address[BRU1]{Service de Physique Th\'eorique, Universit\'e Libre de Bruxelles, B-1050, Brussels, Belgium}
\address[DAV]{University of California Davis Department of Physics, 95616, USA}
\address[BRU2]{Vrije Universiteit Brussel - IIHE, Brussels, Belgium}
\address[NIKHEF]{NIKHEF, Amsterdam, 1098 XG, Netherlands}
\address[ROME]{Universita' di Roma ``Sapienza'' / INFN, Rome, 00185, Italy} 
\address[FREI]{Albert-Ludwigs-Universitaet Physikalisches Institut, Freiburg, 79104, Germany}
\address[GEN]{Universit\'e de Gen\`eve Ecole de Physique, Geneva, CH-1211 Switzerland} 
\address[BONN]{University of Bonn Physikalisches Institut, 53115, Germany} 
\address[KCL]{King's College London Department of Physics, London, WC2R 2LS, United Kingdom}
\address[MAR]{University of Maryland Department of Physics, College Park, 20742-4111, USA} 
\address[OXF]{Rudolf Peierls Centre for Theoretical Physics, University of Oxford, OX1 3NP Oxford, United Kingdom}
\address[ARG]{Physics Division, Argonne National Laboratory, Lemont, 60439, USA}
\address[FNAL]{Fermi National Accelerator Laboratory, Batavia, 60510-5011, USA}
\address[LPSC]{LPSC, Universit\'e Grenoble-Alpes, CNRS/IN2P3, 38042, France} 
\address[UCSB]{University of California Santa Barbara Department of Physics, Santa Barbara, 93106, USA}
\address[MAL]{National Centre for Particle Physics, University of Malaya, Kuala Lumpur, 50603 Malaysia}
\address[ORE]{University of Oregon Department of Physics, Eugene, 97403, USA}
\address[HAR]{Harvard University Department of Physics, Cambridge, 02138, USA} 
\address[DESY]{DESY, Notkestrasse 85, D-22607 Hamburg, Germany}
\address[RHUL]{Royal Holloway University of London Department of Physics, Egham, TW20 0EX, United Kingdom}
\address[UCL]{University College London, WC1E 6BT, United Kingdom}
\address[OSU]{The Ohio State University, Columbus, 43210, USA}
\address[TTU]{Texas Tech University, Lubbock, 41051, USA}
\address[BRO]{Physics Department, Brown University, Providence, 02912, USA}
\address[GRAP]{GRAPPA, University of Amsterdam, 1098 XH, Netherlands}
\address[CHI]{Enrico Fermi Institute, University of Chicago, 60637, USA}
\address[KAV]{Kavli Institute for Cosmological Physics and the Enrico Fermi Institute, The University of Chicago, 60637, USA}
\address[RAL]{Particle Physics Department, Rutherford Appleton Laboratory, OX11 0QX, United Kingdom}
\address[UCI]{Department of Physics and Astronomy, University of California, Irvine, 92697-4575, USA}
\address[KEK]{KEK, Tsukuba, 305-0801, Japan} 
\address[GLA]{University of Glasgow, G12 8QQ, United Kingdom}
\address[BERK]{Berkeley Center for Theoretical Physics, University of California, Berkeley, 94720-7300, USA}
\address[LBNL]{Theoretical Physics Group, Lawrence Berkeley National Laboratory, Berkeley, 94720-8162, USA}
\address[WUP]{Bergische Universitaet Wuppertal D-42119, Germany}
\address[ZUR]{University of Zurich Physik-Institut, CH-8057, Switzerland}
\address[PAV]{INFN Sezione di Pavia, 27100, Italy}
\address[TOR]{University of Toronto Department of Physics, ON M5S 1A7, Canada}
\address[OK]{University of Oklahoma Department of Physics, Norman, 73019, USA}
\address[HEI]{Max-Planck-Institut f\"ur Kernphysik, Heidelberg, 69117, Germany}
\address[REH]{Weizmann Institute of Science, Department of Particle Physics and Astrophysics, Rehovot, 7610001, Israel}
\address[NYU]{New York University Department of Physics, New York, 10003, USA}
\address[PI]{Perimeter Institute for Theoretical Physics, Waterloo, ON N2L 2Y5, Canada}
\address[MCM]{McMaster University Department of Physics \& Astronomy, Hamilton, ON L8S 4M1, Canada}


\begin{abstract}
This document 
 outlines a set of simplified models for dark matter and its interactions with 
Standard Model particles.  It is intended to summarize the main characteristics 
that these simplified models have when applied to dark matter searches at the 
LHC, and to provide a number of useful expressions for reference. The list of 
models includes both $s$-channel and $t$-channel scenarios. For $s$-channel, 
spin-0 and spin-1 mediation is discussed, and also realizations where the Higgs 
particle provides a portal between the dark and visible sectors. The guiding 
principles underpinning the proposed simplified models are spelled out, and 
some suggestions for implementation are presented.
\end{abstract}


\end{frontmatter} 

\vspace{2mm}

\noindent
{\em Preprint:} {\tt CERN-PH-TH/2015-139, FERMILAB-PUB-15-283-CD}\\
{\em Submitted:} June 9, 2015 (revised \& accepted August 3, 2015)\\
{\em Published:} Physics of the Dark Universe, Volumes 9-10, pp 8-23 \\
{\em DOI:} http://dx.doi.org/10.1016/j.dark.2015.08.001\\

\newpage

\section{Introduction}
\label{sec:intro}

Gravitational effects on astrophysical scales give convincing evidence 
for the presence of dark matter (DM) in Nature, an observation that is 
strongly supported by the large-scale structure of the Universe and 
measurements of the cosmic microwave background \cite{Bertone:2004pz}. 
While the existence of 
DM thus seems well established, very little is known about the 
properties of the DM particle(s). To shed light on this question, three 
classes of  search strategies are being employed: $(i)$ direct detection 
in shielded underground detectors; $(ii)$ indirect detection with 
satellites, balloons, and ground-based telescopes looking for signals of 
DM annihilation;  $(iii)$ particle colliders aiming at direct DM 
production. Despite this intense effort, DM  has so far proven elusive. 
In the coming years, direct and indirect detection will 
reach new levels of sensitivity, and the LHC will be operating at 
13~TeV centre-of-mass energy after a very successful~8~TeV run. These 
upcoming experiments will provide crucial tests of our ideas about DM, 
and have great potential to revolutionize our understanding of its 
nature. 

Dedicated searches for DM candidates represent an integral part of the 
physics programme at the LHC.  The minimal experimental signature of DM 
production at a hadron collider consists of an excess of events with a 
single final-state object  $X$ recoiling against large amounts of 
missing transverse momentum or energy ($\slashed{E}_T$).  In Run I of 
the LHC, the ATLAS and CMS collaborations have examined a variety  of 
such ``mono-$X$'' signatures involving  jets of hadrons, gauge bosons, 
top and bottom quarks as well as the Higgs boson in the final state. A 
second class of $\slashed{E}_T$ signatures that has been studied in 
depth arises from the production of ``partner'' particles that decay to 
DM and Standard Model (SM) particles, which usually leads to rather  complex final 
states (for a review of the experimental status after LHC 
Run~I, see for instance \cite{Askew:2014kqa}).

In order to interpret the cross section limits  obtained  from the LHC 
$\slashed{E}_T$ searches, and to relate these bounds to the constraints 
that derive  from direct and indirect detection, one needs a theory of 
DM. In fact, as illustrated in Figure~\ref{fig:simplifiedmodels}, one 
can construct not just one, but a large number of qualitatively 
different DM models. Collectively these models populate the ``theory 
space'' of all possible realizations of physics beyond the SM 
with a particle that is a viable DM candidate. The members 
of this theory space fall into three distinct classes:
\begin{figure}[!t]
\begin{center}
\includegraphics[width=0.8\textwidth]{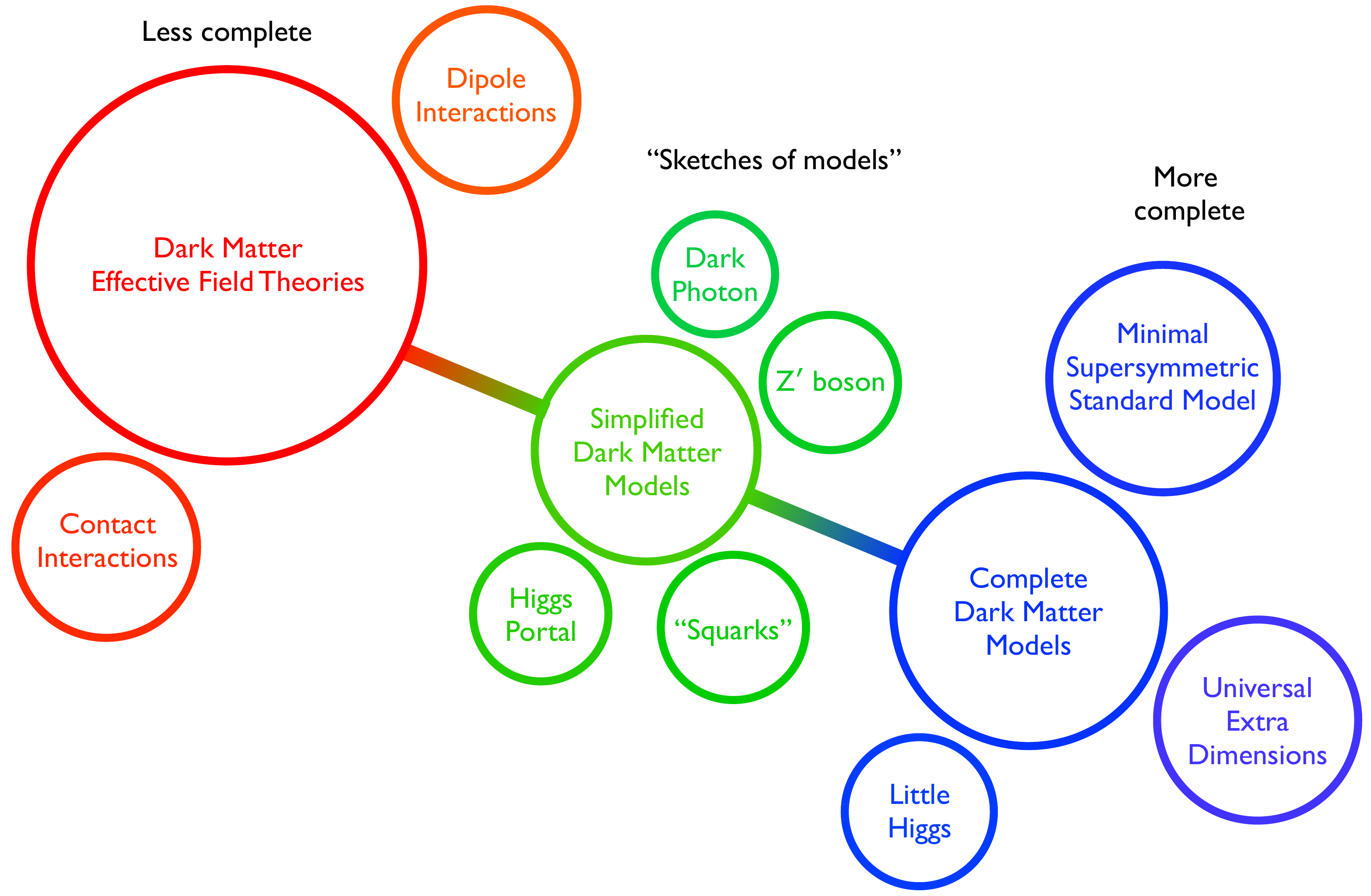} 
\vspace{2mm}
\caption{\label{fig:simplifiedmodels} Artistic view of the DM theory 
space. See text for detailed explanations.}
\end{center}
\end{figure}

\begin{itemize}

\item[(I)] On the simple end of the spectrum, we have theories where 
the DM may be the only  accessible state to our experiments. In such a 
case, effective field theory (EFT) allows us to describe the DM-SM 
interactions mediated by all kinematically inaccessible particles in 
a universal way. The DM-EFT approach \cite{Cao:2009uw,Beltran:2010ww,Goodman:2010yf,Bai:2010hh,Goodman:2010ku,Goodman:2010qn,Rajaraman:2011wf,Fox:2011pm}  
has proven to be very useful in the analysis of LHC Run I data, because 
it allows to derive stringent bounds on the ``new-physics" scale 
$\Lambda$ that suppresses the higher-dimensional operators. Since for 
each  operator a single parameter  encodes the information on all the 
heavy states of the dark sector,  comparing LHC bounds to the limits 
following from direct and indirect DM searches is straightforward in 
the context of DM-EFTs. 

\item[(II)] The large energies accessible at the LHC call into question 
the momentum expansion underlying the EFT approximation~\cite{Bai:2010hh,Fox:2011pm,Fox:2011fx,Shoemaker:2011vi,Busoni:2013lha,Buchmueller:2013dya,Busoni:2014sya,Busoni:2014haa,Racco:2015dxa}, 
and we can expand our level of detail toward simplified DM models (for 
early proposals see for example \cite{Dudas:2009uq,Goodman:2011jq,An:2012va,Frandsen:2012rk,Dreiner:2013vla,Cotta:2013jna}). 
Such models are characterized by the most important state mediating the 
DM particle interactions with the SM, as well as the DM particle itself.
%
%
%
Unlike the 
DM-EFTs,  simplified models are able to describe correctly the full 
kinematics of DM production at the LHC, because they resolve the EFT 
contact interactions  into single-particle $s$-channel or $t$-channel 
exchanges. This comes with the price that they typically involve not 
just one, but a handful of parameters that characterize the dark sector and 
its coupling to the visible sector.  

\item[(III)]  While simplified models capture some set of signals 
accurately at LHC energies (and beyond), they are likely to miss 
important correlations between observables. Complete DM models close 
this gap by adding more particles to the SM, most of which are not 
suitable DM candidates. The classical example is the Minimal 
Supersymmetric SM (MSSM), in which each SM particle gets its own 
superpartner and the DM candidate, the neutralino, is a weakly 
interacting massive particle. Reasonable phenomenological models 
in this class have of order 20 parameters, leading to varied visions of 
DM. 
At the same time, they build-in correlations from symmetry-enforcing 
relations among couplings, that would look like random accidents in a 
simplified model description.
Complete DM models can in principle answer any question satisfactorily, 
but one might worry that their structure is so rich that it is impossible 
to determine unambiguously 
the underlying new dynamics  from a finite amount of data 
(``inverse problem") \cite{ArkaniHamed:2005px}. 

\end{itemize}

 Given our ignorance of the portal(s) between  the dark sector and the 
SM, it is important that we explore {\it all} possibilities that the 
DM theory space has to offer. While the three frameworks discussed 
above have their own {\it pros} and {\it cons},  they are all 
well-motivated, interesting, and 
each could, on its own, 
very well lead to 
breakthroughs in our understanding of DM. Ignoring whole ``continents"  
of the DM theory landscape at Run~II, say EFTs, would be shortsighted, 
and might well make it impossible to exploit the full LHC potential 
as a DM discovery machine. 
 
In recent years, 
a lot of progress has been made in exploring and 
understanding both DM-EFTs and a variety of complete models. The same 
cannot (yet) be said about simplified models that bridge between the 
two ends of the spectrum in theory space. Following the spirit of~\cite{Abdallah:2014hon, Malik:2014ggr},  
we attempt in this document to lay the theoretical groundwork that
should be useful for 
the DM@LHC practitioner. We begin in 
Section \ref{sec:couplings} by discussing the general criteria that a 
simplified DM model should fulfill to make it useful at the LHC. This 
section contains in addition an explanation of the concept of Minimal 
Flavor Violation (MFV) \cite{Hall:1990ac,Chivukula:1987py,Buras:2000dm,D'Ambrosio:2002ex} 
and its importance to model building as well as a brief note on the 
relevance of the spin of the DM particle for LHC searches. Simplified spin-0 
$s$-channel models are then described in Section~\ref{sec:scalar}. 
Since these scenarios can be understood as limiting cases of Higgs 
portal models, we provide in Section~\ref{sec:higgsportal} 
a summary of the most important representatives of these theories.
Section~\ref{sec:vector} is devoted to  simplified spin-1 $s$-channel 
models, while  Section~\ref{sec:tchannel} deals with $t$-channel 
scenarios. To make the work self-contained, we not only discuss the 
LHC phenomenology of each simplified model, but also provide the 
relevant formulae to analyze the constraints from direct detection and 
annihilation of DM.  We conclude and provide an outlook in 
Section~\ref{sec:conclusions}.

 \section{Criteria for Simplified Models}
 \label{sec:couplings}
 
For a simplified DM model to be useful at the LHC, it should fulfill the 
following three criteria: $(i)$ it should be simple enough to form a 
credible unit within a more complicated model; $(ii)$ it should be complete 
enough  to be able to describe accurately the relevant physics phenomena 
at the energies that can be probed at the LHC; $(iii)$  by construction it should  
satisfy all non high-$p_T$ constraints in most of its parameter space.
 
\begin{figure}[!t]
\begin{center}
\includegraphics[width=0.9\textwidth]{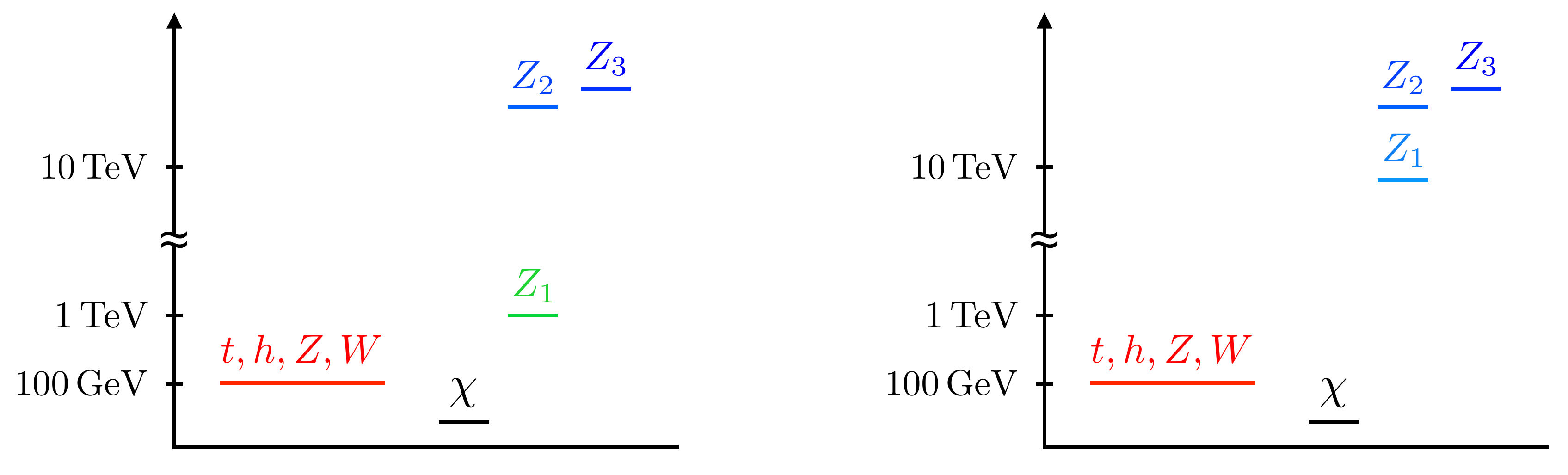} 
\vspace{2mm}
\caption{\label{fig:toyspectrum}  Left: Schematic mass spectrum of a 
simplified DM model. In the case considered, the DM particle $\chi$ is 
lighter than the heaviest SM particles $t,h,Z,W$. The lightest mediator 
state is called $Z_1$ and can be produced on-shell at the LHC. The 
remaining dark-sector states $Z_2$ and $Z_3$ are separated by a mass gap 
from $Z_1$ and inaccessible. Right: The EFT limit of the simplified 
model with a decoupled mediator $Z_1$. See text for further details.}
\end{center}
\end{figure}

One way to guarantee that these three criteria are met consists in putting 
the following requirements/restrictions on the particle content and the 
interactions of the simplified model:
\begin{itemize}
\item[(I)] Besides the SM, the model should contain a DM candidate that 
is either absolutely stable or  lives long enough to escape the LHC 
detectors, as well as a mediator that couples the two sectors.  The dark 
sector can be richer, but the additional states should be somewhat 
decoupled. A typical  mass spectrum is sketched on the left in 
Figure~\ref{fig:toyspectrum}. 

\item[(II)] The Lagrangian  should contain (in principle) all terms that 
are renormalizable and consistent with Lorentz invariance, the  SM gauge 
symmetries, and DM stability. However, it may be permissible to neglect 
interactions or to study cases where couplings are set equal to one 
another. If such simplifications are made, one should however try to 
verify  that these approximations do not result in a very different 
DM phenomenology and they should be spelled out clearly in the 
text 
and on all relevant plots.
 
\item[(III)] The additional interactions should not violate the exact 
and approximate accidental global symmetries of the SM. This means that  
the interactions between the visible and the dark sector should be such  
that  baryon and lepton number is conserved and that the custodial and 
flavor symmetries of the SM are not strongly  broken.   
\end{itemize}

Simplified models are thus specifically designed to involve only a few 
new particles and interactions, and many of them can be understood as 
a limit of a more general new-physics scenario, where all but the 
lightest dark-sector states are integrated out. By construction, the 
physics of simplified models can therefore be characterized in terms 
of a small number of parameters such as particle masses and couplings. 
While simplified models are clearly not model-independent, they do 
avoid some pitfalls of DM-EFTs. In particular, 
they allow one to correctly describe the kinematics of DM production at 
the LHC, by virtue of the dynamical mediator(s) that they contain. 
Conversely, by making the mediator(s) sufficiently heavy the EFT 
framework can be recovered. The latter feature is illustrated on the 
left-hand side of Figure~\ref{fig:toyspectrum}.

\subsection{Note about Flavor and CP Violation}

The requirement (III) deserves further explanations. The  SM  posseses 
both exact and approximate global accidental symmetries. The former 
(baryon and lepton number) are conserved at the renormalizable level, 
while the latter (custodial and flavor symmetries) are broken  by 
quantum effects, but parametrically small in the sense that they become 
exact global symmetries when a parameter or a number of parameters are set 
to zero. New physics will generically not respect these accidental 
symmetries and, as a result, its
parameter space will be severely 
constrained: the new interactions are required to be weak or  the new 
states have to be  heavy (or both).  

A systematic way to curb the size of dangerous flavor-violating  and 
CP-violating effects consists in imposing MFV. 
Loosely speaking the idea behind MFV is that the general structure of 
flavor-changing neutral current (FCNC) processes present in the SM is 
preserved  by new physics. In particular, all flavor-violating and 
CP-violating transitions are governed by the Cabibbo-Kobayashi-Maskawa  
(CKM) matrix.  This basic idea can be formalized and formulated in an 
EFT  \cite{D'Ambrosio:2002ex}. Employing the EFT language, a new-physics 
model satisfies the MFV criterion if the additional interactions  in 
the quark sector are either invariant under the global SM flavor group 
$G_q = U(3)_q \times U(3)_u \times U(3)_d$, or any breaking is 
associated with the quark Yukawa matrices $Y^u$ and $Y^d$.  The notion 
of MFV can be also be extended to the case of  CP violation and to the 
lepton sector --- although for leptons its definition is not unique, 
if one wants to accommodate neutrino masses. 

\subsubsection{MFV Spin-0 $s$-Channel Models}

To understand which restrictions MFV imposes on the flavor structure of 
simplified models, we work out some examples relevant for the discussions 
in later sections. We begin with a very simple model in which DM is a 
real scalar (gauge and flavor) singlet $\chi$ and the SM Higgs doublet 
$H$ provides a portal to the dark sector of the form $\chi^2 |H|^2$ (the 
most important phenomenological implications of  this scenario will be 
discussed in Section \ref{sec:higgsmix}). Following the notion of MFV, 
the interaction terms between the mediator and the quark fields should 
be either invariant under $G_q$ or break it only via $Y^u$ or $Y^d$. 
Given the transformation properties  $q \sim (3, 1, 1)$, $u \sim (1, 3, 1)$ 
and $d \sim (1, 1, 3)$, it follows that the combination $\bar q u$ of 
left-handed and right-handed quark fields  breaks $U(3)_q\times U(3)_u$, 
while the bilinear $\bar q d$  breaks $U(3)_q\times U(3)_d$. This means 
that we have to go with the second option. In terms of gauge eigenstates, 
we write
\beq \label{eq:example1}
{\cal L} \supset -  \sum_{i,j} \Big ( (Y^u)_{ij} \bar q_i H u_j + (Y^d)_{ij} \bar q_i \tilde H d_j + {\rm h.c.} \Big )  \,,
\eeq
where  $i,j$ runs over the three quark families,  
$\tilde H_a = \epsilon_{ab} H^b$ with $a,b=1,2$ and the two terms involve 
the Higgs fields to make them  $SU(2)_L \times U(1)_Y$ gauge invariant. 
Notice that the above interactions are invariant under $G_q$, if the 
Yukawa matrices are promoted to non-dynamical fields (spurions)  with 
the following transformation properties $Y^u \sim (3,\bar 3, 1)$ and 
$Y^d \sim (3, 1 , \bar 3)$. 

Having constructed the couplings between the mediator and the quarks  
in the gauge basis, one still has to transform to the mass eigenstate 
basis. In the case of (\ref{eq:example1}) the final result of this 
transformation is obvious, because the  Lagrangian  is  simply the quark 
part of the Yukawa sector of the SM. One finds 
\begin{equation}
{\cal L} \supset -\frac{h}{\sqrt{2}} \sum_i  \Big ( y_i^u \bar u_i u_i + y_i^d \bar d_i d_i  \Big ) \,,
\end{equation}
where $h$ is the physical Higgs field and $y_i^q = \sqrt{2} m_i^q/v$ with 
$v \simeq 246 \, {\rm GeV}$ the vacuum expectation value (VEV) of $H$ 
that breaks the electroweak symmetry. The lesson to learn from this 
exercise is that in order to construct MFV simplified models that 
describe $s$-channel exchange of spin-0 resonances, the portal couplings 
to the SM fermions should be of Yukawa type. The above line of reasoning 
will be applied to the simplified models in Section~\ref{sec:scalar}. 

\subsubsection{MFV Spin-1 $s$-Channel Models}

The second example that we want to discuss is even simpler than the first 
one. We consider the interactions of DM in form of a Dirac fermion $\chi$  
with the SM quarks through the exchange of spin-1 mediators which we call 
$Z^\prime$. MFV does not restrict the couplings between the mediator and 
DM, and as a consequence the interactions take the  generic form  
$Z^\prime_\mu \,  \big (g_L^{\chi}   \bar \chi \gamma^\mu P_L \chi + g_R^{\chi}   \bar \chi \gamma^\mu P_R \chi \big )$ with $P_{L,R} = (1\mp\gamma_5)/2$ 
denoting left-handed and right-handed chiral projectors. Since the 
bilinears $\bar q \gamma^\mu q$, $\bar u \gamma^\mu u$, and $\bar d \gamma^\mu d$ 
are all flavor singlets, we do not have to invoke the Yukawa couplings $Y^u$ 
and $Y^d$, and simply write 
\beq \label{eq:Lexample2}
{\cal L} \supset Z_\mu^\prime \sum_i \Big [ g_{L}^q  \left ( \bar u_i \gamma^\mu P_L u_i +  \bar d_i \gamma^\mu P_L d_i \right )  + g_R^u \bar u_i \gamma^\mu P_R u_i  + g_R^d \bar d_i \gamma^\mu P_R d_i \Big ] \,.
\eeq
In fact, this expression holds both in the gauge as well as the mass 
eigenstate basis as long as the coefficients $g_{L}^q$, $g_{R}^u$, and 
$g_{R}^d$  are flavor independent. Notice that (\ref{eq:Lexample2}) contains 
the case of pure vector or axialvector quark couplings as a special 
case,~i.e.~$g_{L}^q = g_{R}^u = g_{R}^d$ or $g_{L}^q = -g_{R}^u = -g_{R}^d$, 
respectively. Spin-1 $s$-channel simplified models of MFV type will be 
discussed in Section~\ref{sec:vector}. 

\subsubsection{Comment on Non-MFV Models}

For the sake of argument let us also consider an example of a simplified 
model that does not conform with MFV. As a toy-model we take a $Z^\prime$ 
boson that couples vectorial to the quark gauge eigenstates, but differently 
to the first, compared to
the second and third generations. We parameterize this 
non-universality by a real parameter $\Delta_V$, and restrict ourselves to 
down-type quarks writing
\beq \label{eq:L3} 
{\cal L} \supset Z_\mu^\prime  \sum_i \left (  g_V + \Delta_V \delta_{i1} \right ) \bar d_i \gamma^\mu d_i \,.
\eeq
To go to the mass eigenstate basis requires rotating the left-handed and 
right-handed quark fields by $3 \times 3$ unitarity matrices $U_{L,R}^{u,d}$.  
These rotations will leave the term  proportional to $g_V$ flavor diagonal, 
but will induce flavor off-diagonal terms of the form 
\begin{equation}  \label{eq:L4} 
{\cal L} \supset Z_\mu^\prime  \hspace{0.25mm} \Delta_V \sum_{i,j} \left ( L_{ij} \bar d_i \gamma^\mu P_L d_j +  R_{ij} \bar d_i \gamma^\mu P_R d_j  \right ) \,.
\end{equation}
with 
\beq
L = U_L^{d \, \dagger} \hspace{0.5mm} {\rm diag} \left (1, 0, 0 \right ) U_L^d \,, \qquad
R = U_R^{d \, \dagger} \hspace{0.5mm} {\rm diag} \left (1, 0, 0 \right ) U_R^d \,.
\eeq

At this point we have to make some assumptions about the flavor structure 
of the ultraviolet~(UV) complete model that gives rise to (\ref{eq:L3}) to 
progress further. Since the right-handed rotations $U_R^{u,d}$ are not 
observable in the SM, we assume that $U_R^d$ is the $3\times 3$ unit 
matrix $1_3$. This implies that $R = {\rm diag} \left (1,0,0 \right)$ and 
thus there are no FCNCs in the right-handed down-quark sector. In contrast, 
the  left-handed rotations are observable in the SM, because  they combine 
to give the CKM matrix,~i.e.~$V = U_L^{u \, \dagger} U_L^d$.  A possible 
simple choice that satisfies this requirement is $U_L^d = V$ and 
$U_L^u = 1_3$, resulting in 
\beq \label{eq:Lmatrix}
L = \begin{pmatrix} |V_{ud}|^2 & V_{ud}^\ast V_{us} & V_{ud}^\ast V_{ub}  \\  V_{us}^\ast V_{ud}  & |V_{us}|^2 & V_{us}^\ast V_{ub}   \\  V_{ub}^\ast V_{ud}  &  V_{ub}^\ast V_{us}  & |V_{ub}|^2 \end{pmatrix}  \,,
\eeq
which implies flavor violation in the down-type quark sector. Note that 
choosing  $U_L^d = 1_3$ and $U_L^u = V^\dagger$ would give 
$L = {\rm diag} \left (1,0,0 \right )$. However, this choice does not solve 
the new-physics ``flavor problem", because it is easy to see that FCNCs 
would then appear in the up-type quark sector. 

Using  (\ref{eq:Lmatrix}) it is straightforward to calculate the FCNCs 
induced by tree-level $Z^\prime$-boson exchange. For instance, the 
new-physics amplitude relevant for kaon mixing can be estimated to be 
\beq \label{eq:Amplitude_zprime}
{\cal A} ( s \bar d \to Z^\prime \to \bar s d ) \sim  \frac{(V_{ud}^\ast V_{us})^2 \Delta_V^2 }{M_{Z^\prime}^2} \simeq 5 \times 10^{-2} \, \frac{\Delta_V^2 }{M_{Z^\prime}^2} \,.
\eeq
with $M_Z^\prime$ the mass of the $Z^\prime$ boson. This result should be 
compared to the dominant SM contribution to $K$--$\bar K$ mixing, which 
arises from top-$W$ boxes and is given by 
\beq 
{\cal A} ( s \bar d \to {\rm box} \to \bar s d ) \sim  \frac{\alpha_w^2 \left (V_{td}^\ast V_{ts} \right)^2 y_t^2}{256 \hspace{0.25mm} M_W^2} \simeq \frac{5 \times 10^{-13} }{M_W^2} \,,
\label{eq:Amplitude_tbox}
\eeq
with $\alpha_w = g^2/(4 \pi)$ the weak coupling constant, $M_W$ the 
$W$-boson mass, and $y_t \simeq 1$ the top Yukawa.  ~A rough bound on the 
amount of additional flavor violation $\Delta_V/M_{Z^\prime}$ can now be 
obtained by simply requiring that (\ref{eq:Amplitude_zprime}) should be 
smaller in magnitude than (\ref{eq:Amplitude_tbox}). It follows that 
\beq \label{eq:flavorproblem}
\left | \frac{\Delta_V M_W}{M_{Z^\prime}} \right | \lesssim 3 \times 10^{-6} \,,  
\eeq
which implies that for $\Delta_V \simeq 1$ the $Z^\prime$-boson mass 
$M_{Z^\prime}$ should be larger than around $3 \times 10^4 \, {\rm TeV}$, 
because otherwise one would be in conflict with  the experimental bounds 
on kaon mixing.   In view of this result it should be clear that in 
order to  allow for interesting LHC phenomenology, one has to require 
that the simplified model is MFV. In our toy model (\ref{eq:L3}) this is 
simply achieved by setting $\Delta_V =0$. We finally add that for very 
light mediators  important constraints on simplified models can however 
still arise from quark-flavor physics even if their interactions are MFV 
(see \cite{Dolan:2014ska} for a recent comprehensive discussion).

\subsection{Note about Spins}
 
In many cases, there will be variations of the simplified model under 
consideration where the DM is a real or complex scalar, Dirac or Majorana 
fermion, or even a neutral vector.  In some cases, even simple changes 
such as considering a Majorana instead of a Dirac fermion can lead to big 
changes in the phenomenology of direct detection experiments and/or 
annihilation. The classical examples are that for Majorana fermions the 
vector coupling vanishes identically and that such DM particles cannot have an electric 
or magnetic dipole moment. In the context of simple cut-and-count analyses 
at the LHC, the precise nature of the DM particle is generically less 
relevant in the sense that it  will to first order only affect the total 
production cross sections. Angular observables that are sensitive to the 
structure of the dark sector have however been constructed and 
studied~\cite{Cotta:2012nj,Haisch:2013fla,Crivellin:2015wva}, but such 
analyses necessarily involve topologies beyond $2 \to  \slashed{E}_T + 1$. 
 
\section{Scalar $\bm s$-channel Mediator}
 \label{sec:scalar}
 
 A scalar particle mediator can be a very simple addition to the 
SM.  If it is chosen as a gauge singlet, it can have 
tree-level interactions with a singlet DM particle that 
is either a Dirac or Majorana fermion, or DM that is 
itself a scalar. The spin-$0$ mediator could still be chosen as 
either a real or complex scalar, which are distinguished by the 
fact that a complex scalar contains both scalar and pseudoscalar 
particles, whereas the real option contains only the scalar field. 
We will consider here two choices for DM simplified models: 
one where the interaction with the SM is mediated by 
the real scalar, and the second where we consider only a light 
pseudoscalar (assuming that the associated scalar is decoupled 
from the low-energy spectrum). 
 
Couplings to the SM fermions can be arranged by mixing 
with the SM Higgs.  Such models have intriguing 
connections with Higgs physics, and can be viewed as 
generalizations of the Higgs portal to DM.  The impact on 
Higgs physics is discussed in Section~\ref{sec:higgsmix} below.  
The most general scalar mediator models will of course have 
renormalizable interactions between the SM Higgs and 
the new scalar $\phi$ or pseudoscalar $a$, as well as $\phi/a$ 
interactions with electroweak gauge bosons. Such interactions are 
model-dependent, often subject to constraints from electroweak 
precision tests, and would suggest specialized searches which 
cannot be generalized to a broad class of models (unlike, for 
instance, the $\missET+ j$ searches). As a result, for this class 
of simplified models with spin-$0$ mediators, we suggest to focus 
primarily on the couplings to fermions and the loop-induced 
couplings to gluons. The possibility that the couplings to the 
electroweak sector can also lead to interesting DM 
phenomenology should however be kept in mind, and can be studied 
in the context of Higgs portal DM.

\subsection{Fermionic DM}

MFV dictates that the coupling of a scalar to the SM 
fermions will be proportional to the fermion masses.  However, it 
allows these couplings to be scaled by separate factors for the 
up-type quarks, down-type quarks, and the charged leptons. 
Assuming that DM is a Dirac fermion $\chi$, which couples to 
the SM only through a scalar $\phi$ or pseudoscalar 
$a$, the most general tree-level Lagrangians compatible with the 
MFV assumption are~\cite{Cotta:2013jna,Abdullah:2014lla}:
 \begin{eqnarray}
{\cal L}_{{\rm fermion},\phi} & \supset & - g_\chi \phi \bar{\chi}\chi - \frac{\phi}{\sqrt{2}} \sum_i \left(g_u y_i^u \bar{u}_i u_i+g_d y_i^d \bar{d}_i d_i+g_\ell y_i^\ell \bar{\ell}_i \ell_i\right)\, , \label{eq:scalarlag} \\
{\cal L}_{{\rm fermion},a} & \supset & - ig_\chi a \bar{\chi}\gamma_5\chi - \frac{i a}{\sqrt{2}}\sum_i  \left(g_u y_i^u \bar{u}_i \gamma_5 u_i+g_d y_i^d \bar{d}_i \gamma_5 d_i+g_\ell y_i^\ell   \bar{\ell}_i \gamma_5 \ell_i\right) \,. \label{eq:pseudoscalarlag}
\end{eqnarray}
Here the sums run over the three 
 SM families and we 
are using Yukawa couplings $y_i^f$ normalized as 
$y_i^f = \sqrt{2}m_i^f/v$ with $v$ the  
Higgs VEV. 
We parametrize the DM-mediator coupling by 
$g_\chi$, rather than by a Yukawa coupling $y_\chi = \sqrt{2}m_\chi/v$, 
since the the DM particle $\chi$ 
most likely receives its mass from other (unknown) mechanisms, rather than 
electroweak symmetry breaking.

The most general Lagrangians including new scalars or 
pseudoscalars will have a potential  containing interactions 
with the SM Higgs field $h$. As stated above, we 
choose to take a more minimal set of possible interactions, and 
leave the discussions of the couplings in the Higgs sector to 
the section on  Higgs portal DM. Given this 
simplification, the minimal set of parameters under 
consideration is
 \bea
  \left\{ m_\chi,~ m_{\phi/a},~ g_\chi,~ g_u,~ g_d,~ g_\ell \right\} \,.
 \eea
The simplest choice of couplings is  $g_u = g_d = g_\ell$, 
which is realized in singlet scalar extensions of the SM (see 
Section~\ref{sec:higgsmix}).  If one extends the SM 
Higgs sector to a two-Higgs-doublet model, one can obtain other 
coupling patterns such as  $g_u \propto \cot \beta$ and 
$g_d \propto g_e \propto \tan \beta$ with $\tan \beta$ denoting the ratio 
of VEVs of the two Higgs doublets.  The case 
$g_u \neq  g_d \neq g_\ell$ requires additional scalars, 
whose masses could be rather heavy. For simplicity, we will 
use universal 
couplings $g_v = g_u = g_d = g_\ell$ in the remainder of this 
section, though one should bear in mind that finding ways to 
test this assumption experimentally would be very useful.

The signal strength in DM pair production does not only 
depend on the masses $m_\chi$ and $m_{\phi/a}$ and the couplings 
$g_i$, but also on the total decay width of the mediator $\phi/a$. 
In the minimal model as specified by \myeq{eq:scalarlag} and 
\myeq{eq:pseudoscalarlag}, the widths for the mediators are given 
by:
\begin{eqnarray}
\Gamma_\phi & = & \sum_f N_c \frac{y_f^2 g_v^2 m_\phi}{16 \pi} \left(1-\frac{4 m_f^2}{m_\phi^2}\right)^{3/2} + \frac{g_\chi^2 m_\phi}{8 \pi} \left(1-\frac{4 m_\chi^2}{m_\phi^2}\right)^{3/2} + \frac{\alpha_s^2 g_v^2 m_\phi^3}{32 \pi^3 v^2} \left| f_\phi\left(\tfrac{4m_t^2}{m_\phi^2} \right)\right|^2 \,,  \label{eq:scalarwidth}  \hspace{8mm} \\
\Gamma_a & = & \sum_f N_c \frac{y_f^2 g_v^2 m_a}{16 \pi} \left(1-\frac{4 m_f^2}{m_a^2}\right)^{1/2} + \frac{g_\chi^2 m_a}{8 \pi} \left(1-\frac{4 m_\chi^2}{m_a^2}\right)^{1/2}  
+ \frac{\alpha_s^2 g_v^2 m_a^3}{32 \pi^3 v^2} \left| f_a\left(\tfrac{4m_t^2}{m_\phi^2} \right)\right|^2 \,, \label{eq:pseudoscalarwidth} 
\end{eqnarray}
with 
\bea \label{eq:fphifa}
f_\phi (\tau) = \tau \left [ 1+ (1-\tau) \arctan^2 \left ( \frac{1}{\sqrt{\tau-1}} \right ) \right ]  \,, \qquad 
f_a (\tau) =  \tau \arctan^2 \left ( \frac{1}{\sqrt{\tau-1}} \right ) \,. 
\eea
The first term in each width corresponds to the decay into 
SM fermions (the sum runs over all kinematically 
accessible
fermions, $N_c = 3$ for quarks and $N_c = 1$ for 
leptons). The second term is the decay into DM 
(assuming that this decay is kinematically allowed). The factor 
of two between the decay into SM  fermions and into DM  
is a result of our choice of normalization of the Yukawa 
couplings. The last term corresponds to decay into gluons.  
Since we have assumed that $g_v = g_u = g_d = g_\ell$, we have 
included in the partial decay widths $\Gamma (\phi/a \to gg)$ 
only the contributions stemming from top loops, which provide 
the by far largest corrections given that $y_t \gg y_b$~etc. 
At the loop level the mediators can decay not only to gluons 
but also to pairs of photons and other final states if these 
are kinematically accessible. The decay rates 
$\Gamma (\phi/a \to gg)$ are however always larger than the 
other loop-induced partial widths, and in consequence the 
total decay widths $\Gamma_{\phi/a}$ are well approximated by 
the corresponding sum of the individual partial decay widths 
involving DM, fermion or gluon pairs. Notice finally 
that if  $m_{\phi/a} > 2m_t$  and $g_u \gtrsim g_\chi$ the total widths of $\phi/a$ will 
typically be dominated by the partial widths to top quarks.

\subsubsection{LHC Searches}

Under the assumption of MFV, supplemented 
by $g_v = g_u = g_d = g_\ell$, the most relevant couplings 
between DM and the SM arising from 
\myeq{eq:scalarlag} and \myeq{eq:pseudoscalarlag} are those 
that involve  top quarks.  Two main strategies have been 
exploited to search for scalar and pseudoscalar interactions 
of this type using LHC data. The first possibility consists in 
looking for a mono-jet plus missing energy signal ${\slashed E}_T + j$, where the mediators that pair produce DM are radiated 
from top-quark loops \cite{Haisch:2012kf}, 
while the second possibility relies on 
detecting the top-quark decay products that arise from the 
tree-level reaction ${\slashed E}_T + t \bar t$ \cite{Lin:2013sca}. 
In the  first  paper~\cite{Haisch:2012kf} that discussed the ${\slashed E}_T + j$ signal the effects of DM fermions coupled to heavy-quark loops were characterized in terms
of effective higher-dimensional operators,~i.e.~with mediators being integrated out. The effects of dynamical scalar and pseudoscalar messengers
in the $s$-channel mediating interactions between the heavy quarks in the loop and DM were
computed in characterizing the LHC signatures for DM searches in~\cite{Fox:2012ru,Haisch:2013fla,Buckley:2014fba, Harris:2014hga,Haisch:2015ioa}.

Final states involving top-quark pairs were considered in 
the articles~\cite{Artoni:2013zba, CMS:2014mxa, CMS:2014pvf, 
Aad:2014vea, Buckley:2014fba, Haisch:2015ioa}. 
Searches for a $\slashed{E}_T + b \bar b$ signal~\cite{Lin:2013sca, 
Artoni:2013zba, Aad:2014vea} also provide an interesting avenue 
to probe \myeq{eq:scalarlag} and  \myeq{eq:pseudoscalarlag}, 
while the constraints from mono-jet searches on the scalar and 
pseudoscalar interactions  involving the light quark flavors 
are very weak  due to the strong Yukawa suppression (as 
discussed in detail in~\cite{Fox:2012ru, Haisch:2013ata}), and 
thus are unlikely to be testable at the LHC.  Scenarios where 
the DM-SM interactions proceed primarily via gluons have also 
been considered \cite{Godbole:2015gma}.

\begin{figure}[!t]
\begin{center}
\includegraphics[width=0.95\textwidth]{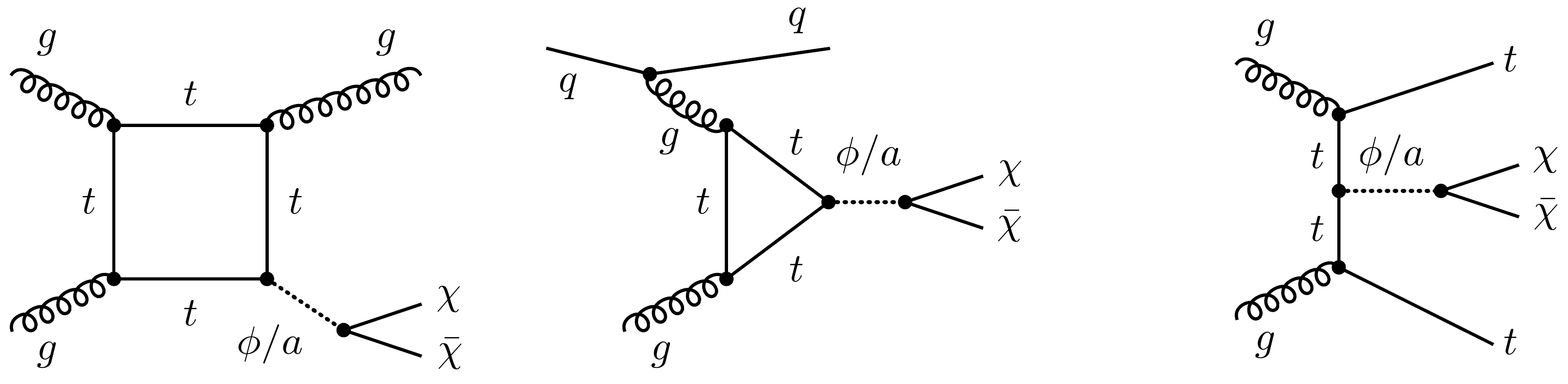} 
\vspace{2mm}
\caption{\label{fig:tops} Left: Two examples of  one-loop  
diagrams with an exchange of a $\phi/a$ mediator that provide 
the dominant contribution to a mono-jet signature. Right: A 
tree-level graph that leads to a $\missET + t \bar t$ signal. }
\end{center}
\end{figure}

Predicting mono-jet cross sections in the simplified models  
\myeq{eq:scalarlag} and  \myeq{eq:pseudoscalarlag} is complicated 
by the fact that the  highly energetic initial-state and/or 
final-state particles involved in the process are able to 
resolve the structure of the top-quark loops that generate the 
$\missET + j$ signal (see the left-hand side of 
Figure~\ref{fig:tops}). Integrating out the top quark and 
describing the 
interactions by an effective operator of 
the form $\phi  G_{\mu \nu}^a G^{a, \mu \nu}$  
($a G_{\mu \nu}^a \tilde G^{a, \mu \nu}$) with $G_{\mu \nu}^a$ 
the field strength tensor of QCD and 
$\tilde G^{a, \mu \nu} = 1/2 \hspace{0.25mm}\epsilon^{\mu \nu \lambda \rho} G^a_{\lambda \rho}$ 
its dual, is in such a situation a poor approximation~\cite{Haisch:2012kf,Fox:2012ru}. 
Already in the LHC Run~I environment the $m_t \to \infty$ limit 
overestimates the exact cross sections by a factor of 
5 (40) for $m_\chi \simeq 10 \,{\rm GeV}$ ($m_\chi \simeq 1 \,{\rm TeV}$)~\cite{Haisch:2015ioa}. 
Removing the top quark as an active degree of freedom becomes 
even less justified at $13 \, (14) \, {\rm TeV}$, where the  
$\slashed{E}_T$ and $p_{T,j}$ selection requirements have to 
be harsher than at $(7) \, 8 \, {\rm TeV}$ to differentiate 
the DM signal from the SM background. In 
order to infer reliable bounds on~\myeq{eq:scalarlag} and  
\myeq{eq:pseudoscalarlag}, one therefore has to calculate the 
mono-jet cross section keeping the full top-quark mass 
dependence. Such 
calculations are now publicly available at leading order~(LO) 
in {\tt MCFM}~\cite{Fox:2012ru} and at LO plus parton shower 
(LOPS) in the {\tt POWHEG~BOX}~\cite{Haisch:2015ioa}. Given 
that the $\missET + t \bar t \, (b \bar b)$ signals arise  in 
the context of   \myeq{eq:scalarlag} and  
\myeq{eq:pseudoscalarlag} at tree level (see the right-hand 
side of Figure~\ref{fig:tops}), event generation through 
programs like {\tt MadGraph5}~\cite{Alwall:2011uj} is possible, 
and UFO model files~\cite{Degrande:2011ua} from different 
groups~\cite{Buckley:2014fba, Harris:2014hga,Haisch:2015ioa} are available for 
this purpose. 

Since \myeq{eq:scalarlag} and \myeq{eq:pseudoscalarlag} is a 
simplified DM model, it is possible that the mediator can decay
into additional states present in the  full theory that we have 
neglected. For example,  $\phi/a$  may decay into new charged 
particles which themselves eventually decay into DM, 
but with additional visible particles that would move the event 
out of the selection criteria of the mono-jet or similar 
$\missET$ searches. 
Another possibility is that the mediator can also decay invisibly
into other particles of the dark sector.
In either case, the expressions for $\Gamma_{\phi/a}$ 
as given in \myeq{eq:scalarwidth} and \myeq{eq:pseudoscalarwidth} 
are lower bounds on the total decay-width of  the mediators.  To 
understand how the actual value of $\Gamma_{\phi/a}$ influences  
the LHC sensitivity, one has to recall that for 
$m_{\phi/a} \ll \sqrt{\hat{s}}$ \, (where $\sqrt{\hat{s}}$ is 
some characteristic fraction of the center-of-mass energy of the 
collider in question) and $m_{\phi/a} > 2 m_\chi$, DM-pair 
production proceeds dominantly via an on-shell mediator. 
If the narrow width approximation (NWA) is applicable, the mono-jet cross section 
factorizes into a product of on-shell production of $\phi/a$ 
times its branching ratio into $\chi \bar \chi$,~i.e.~$\sigma (pp \to \missET + j) = \sigma (pp \to \phi/a + j) \, {\rm Br} (\phi/a \to \chi \bar \chi)$. 
One can draw three conclusions from this result. First, in the 
parameter region where $m_{\phi/a} > 2 m_\chi$ and $\Gamma_{\phi/a} \ll m_{\phi/a}$, 
a change in $\Gamma_{\phi/a}$ will simply lead to a rescaling 
of the cross section, namely $\sigma (pp \to \missET + j) \propto 1/\Gamma_{\phi/a}$. 
This implies in particular that kinematic distributions of simple 
$\missET$ signals will  to first approximation be unaltered under 
variations of $\Gamma_{\phi/a}$. Second, for parameter choices 
where the partial decay width to $\bar{\chi}\chi$ DM pairs is dominant, 
the cross section scales as $\sigma (pp \to \missET + j) \propto g_v^2$. 
If the partial decay width to SM particles gives 
the largest contribution to $\Gamma_{\phi/a}$, one has instead 
$\sigma (pp \to \missET + j) \propto g_\chi^2$. Third, the 
scaling $\sigma (pp \to \missET + j) \propto g_\chi^2 g_v^2$ 
only holds for off-shell production, which  occurs  for 
$m_{\phi/a} < 2 m_\chi$. Notice that for $m_{\phi/a} \lesssim 2 m_\chi$,  
the total decay width of $\phi/a$ will have a non-trivial impact 
on the constraints that the LHC can set, since the amount of 
off-shell production depends sensitively on $\Gamma_{\phi/a}$. 

Similarly, the total decay width effect is non-trivial when the mediator 
can decay into other particles in the invisible sector beyond the cosmologically stable DM.
To apply the simplified models framework to these scenarios, it was proposed in \cite{Buckley:2014fba,Harris:2014hga} to treat the mediator width 
as an independent parameter in the simplified model characterization.

We now turn to the constraints on these models from non-collider 
experiments: thermal relic abundance, indirect detection, and 
direct detection. The first two results can be considered together, 
as they depend on the same set of annihilation cross sections.

\subsubsection{Thermal Cross Sections}

The thermally-average annihilation of DM through the 
spin-0 mediators can be calculated from the simplified model 
\myeq{eq:scalarlag} and  \myeq{eq:pseudoscalarlag}. The resulting 
cross sections for annihilation into SM fermions are given by 
\begin{eqnarray}
( \sigma v) (\chi\bar{\chi} \to \phi \to f\bar{f}) & = & N_c \, \frac{3 g_\chi^2 g_v^2 y_f^2 \hspace{0.5mm} m_\chi T}{8\pi \left[ (m_\phi^2 - 4m_\chi^2)^2 + m_\phi^2\Gamma^2_\phi \right]}   \left ( 1-\frac{m_f^2}{m_\chi^2} \right )^{3/2} \, , \label{eq:scalarthermal} \\
(\sigma v ) (\chi\bar{\chi} \to a \to f\bar{f}) & = & N_c \, \frac{g_\chi^2 g_v^2 y_f^2 \hspace{0.5mm} m_\chi^2  }{4\pi \left[ (m_a^2 - 4m_\chi^2)^2 + m_a^2\Gamma^2_a \right]}  \left ( 1-\frac{m_f^2}{m_\chi^2} \right )^{1/2} \,,  \label{eq:pseudoscalarthermal} 
\end{eqnarray}
where $T$ denotes the DM temperature.  Notably, scalar mediators do not have a temperature-independent 
contribution to their annihilation cross section, while pseudoscalars 
do.  As $T \propto v^2$ (where $v$ is the DM velocity), there 
is no velocity-independent annihilation through scalars.  In the 
Universe today $v \simeq 1.3 \times 10^{-3} \hspace{0.25mm} c$, so 
there are no non-trivial constraints on DM annihilation from 
indirect detection in the scalar mediator model (see, however, references \cite{Abramowski:2012apj,Aharonian:2008wt}).

The parameter space of the pseudoscalar model, on the other hand, can 
be constrained by indirect detection. Most constraints from indirect 
detection are written in terms of a single annihilation channel, and 
so the constraints for the full simplified model (with multiple 
annihilation channels open) require some minor modifications of the 
available results. In the case at hand, good estimates can be obtained 
by considering the most massive fermion into which the DM can 
annihilate (bottom and top quarks if they are accessible), as they 
dominate the annihilation cross section. Note that, away from 
resonance, the total width $\Gamma_a$ entering in 
\myeq{eq:pseudoscalarthermal} is relatively unimportant for obtaining the 
correct indirect detection constraints.

The thermal relic calculation requires the same input cross sections as  
indirect detection. Here, the cross sections are summed over all 
kinematically available final states, and can be written as
\beq
\langle \sigma v\rangle = a + bT \,.
\eeq
If the DM particles are Dirac fermions, one has to include a 
factor of $1/2$ in the averaging, because Dirac fermions are not their 
own anti-particles. In the Majorana case no such factor needs to be 
taken into account. The thermal relic abundance of DM is then 
\begin{equation}
\Omega_\chi h^2 = 0.11 \, \frac{7.88\times 10^{-11} x_f \, {\rm GeV}^{-2}}{a + 3b/x_f} \,,
\end{equation}
where $x_f = m_\chi/T_f  \in [20,30]$ with $T_f$ the freeze-out 
temperature. For reasonable early Universe parameters, the correct 
relic abundance $\Omega_\chi h^2  \simeq 0.11$ occurs in the ballpark of 
\begin{equation}
3 \times 10^{-26}~\mbox{cm$^3$/s} = 2.57 \times 10^{-9}~\mbox{GeV}^{-2} = a + 3b/x_f \,.
\end{equation}
Keep in mind that these equations require some modification when the 
DM-mediator system is on resonance. Further, recall that it 
is unknown whether or not DM is a thermal relic, or if the 
only annihilation process in play in the early Universe proceeds 
through the mediator considered in the simplified model. Therefore, 
while it is appropriate to compare the sensitivity of experimental 
results to the thermal cross section, this is not the only range of 
parameters of theoretical interest.

\subsubsection{Direct Detection}

In contrast to the 
situation 
discussed before, elastic scattering of DM on nucleons induced 
by $\phi/a$ exchange can be very well described in terms of an EFT.  
Integrating out the mediators leads to the expressions 
\beq \label{eq:Qphia}
O_\phi = \frac{g_\chi g_v y_q}{\sqrt{2} m_\phi^2}\,  \bar \chi \chi \bar q q \,, \qquad  O_a = \frac{g_\chi g_v y_q}{\sqrt{2} m_a^2} \, \bar \chi \gamma_5 \chi \bar q  \gamma_5 q \,,
\eeq
at tree level, as well as contact terms consisting of four DM or 
quark fields. Removing the top quark as an active degree of freedom 
generates an effective interaction between DM and gluons. At the 
one-loop level, one obtains
\beq \label{eq:SVZ}
O_G = \frac{\alpha_s \hspace{0.25mm} g_\chi g_v }{12 \pi v m_\phi^2} \, \bar \chi \chi G_{\mu \nu}^a G^{a, \mu \nu} \,, \qquad O_{\tilde G} =  \frac{\alpha_s \hspace{0.25mm} g_\chi g_v }{8 \pi v m_a^2}  \, \bar \chi \gamma_5 \chi G_{\mu \nu}^a \tilde G^{a, \mu \nu} \,,
 \eeq
by employing the Shifman-Vainshtein-Zakharov relations 
\cite{Shifman:1978zn}.  At the bottom- and  charm-quark threshold, 
one has to integrate out the corresponding heavy quark by again 
applying \myeq{eq:SVZ}. Note that this matching procedure is 
crucial to obtain the correct DM-nucleon scattering cross section 
associated with effective spin-0 DM-quark interactions.

The DM scattering cross section with nuclei is then 
obtained by calculating the nucleon matrix elements of the 
operators \myeq{eq:Qphia} and \myeq{eq:SVZ} at a hadronic scale 
of the order of $1 \, {\rm GeV}$. 
Direct detection provides relevant constraints only on the 
scalar mediator model and not the pseudoscalar case, since 
only the operators $O_\phi$ and $O_G$ lead to a spin-independent 
(SI) cross section, while for $O_a$ and $O_{\tilde G}$ the 
DM-nucleon scattering turns out to be spin-dependent  (SD) and 
momentum-suppressed.

The scalar interactions 
with the nuclear targets used for direct detection 
are (to good approximation) isospin-conserving, so that the 
elastic DM-nucleon cross section can be written as 
($N =n,p$)
\beq \label{eq:sigmaDMN}
\sigma_{\chi-N}^{\rm SI} = \frac{\mu_{\chi-N}^2 m_N^2}{\pi} \left ( \frac{g_\chi g_v}{v m_\phi^2} \right )^2 \, f_N^2 \,,
\eeq
where $\mu_{\chi-N}$ is the DM-nucleon reduced mass 
$\mu_{\chi-N} = m_\chi m_N/(m_\chi + m_N)$ and 
$m_N \simeq 0.939 \, {\rm GeV}$ is the average nucleon mass. 
The form factor $f_N$ is given by 
\beq
f_N = \sum_{q=u,d,s} f_N^q + \frac{2}{27} \hspace{0.25mm} f_N^G \, \simeq \, 0.2 \,,
\eeq
where the numerical value has been obtained using 
$f_N^u \simeq0.017$, $f_N^d \simeq0.036$, $f_N^s  \simeq0.043$ 
\cite{Crivellin:2013ipa,Junnarkar:2013ac} and $f_N^G = 1- \sum_{q=u,d,s} f_N^q  \simeq 0.904$. 
Notice that the constraints arising from existing and future 
direct limits on \myeq{eq:sigmaDMN} can be evaded by assuming 
that $\chi$ is not stable on cosmological time scales, but 
lives long enough to escape the ATLAS and CMS detectors. When 
comparing  the bounds set by direct detection and the LHC, this 
loophole should be kept in mind. 

\section{Higgs Portal DM}
\label{sec:higgsportal}

DM may predominantly couple to the SM particles through the SM Higgs.  There are three broad classes of models of this kind:
\begin{itemize}
\item[A.] The DM particle is a scalar singlet under the SM gauge group, 
which couples through a quartic interaction with the Higgs. The collider phenomenology of this DM scenario has been extensively studied in the literature~(see for instance \cite{Burgess:2000yq,Barger:2007im,Djouadi:2011aa,Djouadi:2012zc,Cline:2013gha,Khoze:2014xha,Craig:2014lda}).

\item[B.] The DM particle is a fermion singlet under the gauge symmetries of the SM, which couples to a scalar boson which  itself mixes with the Higgs.  This model class provides a specific realization of the $s$-channel scalar mediator case discussed in Section~\ref{sec:scalar}.  Its implications for the LHC have been studied for example in \cite{Kim:2008pp,Baek:2011aa,LopezHonorez:2012kv,Carpenter:2013xra}. 

\item[C.] The DM particle itself may be a mixture of an electroweak singlet and doublet \cite{Enberg:2007rp,Mahbubani:2005pt,Cheung:2013dua}, as in the MSSM where it has both bino and higgsino components.  Generically, this is  referred to as ``singlet-doublet" DM \cite{Cohen:2011ec}. 
\end{itemize}
The first two cases capture important features of models \cite{Hambye:2013dgv,Khoze:2014xha,Altmannshofer:2014vra} where the SM is extended to be classically scale invariant \cite{Coleman:1973jx, Meissner:2006zh, Foot:2007iy,Englert:2013gz} with the aim of addressing the electroweak gauge hierarchy problem.

\subsection{Scalar Singlet DM}

In the case where an additional real scalar singlet $\chi$ is the DM candidate, the Lagrangian of the scalar Higgs portal can be written as 
\begin{align} \label{eq:LHP1}
\mathcal{L}_\text{scalar,$H$} \supset  - \lambda_\chi  \chi^4  - \lambda_p \chi^2 |H|^2 \, ,
\end{align}
where $H$ denotes the usual SM Higgs doublet. Augmenting the Lagrangian with  a discrete $Z_2$ symmetry that takes $\chi \rightarrow -\chi$ and  $H \rightarrow H$  leads to stable DM, and in addition guarantees that there is no singlet-Higgs mixing, which leaves the couplings of the SM Higgs unaltered at tree level.  The self-coupling $\lambda_\chi$ of the scalar $\chi$ is in general irrelevant to determining how well the portal coupling $\lambda_p$ can be probed  through LHC DM searches, and thus may be ignored. 

For $m_h > 2 m_\chi$, the most obvious manifestation of the interactions (\ref{eq:LHP1}) is through their contributions to the invisible decay  of the Higgs. The corresponding decay width reads 
\begin{align}
\Gamma (h \to \chi \chi) = \frac{\lambda_p^2  v^2}{2 \pi m_h} \left ( 1 - \frac{4 m_\chi^2}{m_h^2} \right)^{1/2} \,,
\end{align}
with $m_h$ the Higgs mass and $v$ its VEV.  In fact, both ATLAS~\cite{Aad:2014iia} and CMS~\cite{Chatrchyan:2014tja} have already interpreted their Run~I $h \to \rm{invisible}$ searches in terms of the Higgs portal scenario (\ref{eq:LHP1}). For DM candidates with $m_\chi \lesssim 10 \, {\rm GeV}$ these searches are competitive with or even stronger than the SI results provided by direct  detection experiments.  

When $m_h < 2 m_\chi$, the Higgs cannot decay on-shell to a pair of $\chi$ particles, so that DM pair production necessarily has to proceed off-shell. The cross section for this process is then suppressed by an additional factor of $\lambda_p^2$ as well as the two-body phase space, leading to a rate that rapidly diminishes with $m_\chi$. This feature makes a LHC discovery  challenging  even at $14 \,{\rm TeV}$ and high luminosity~\cite{Craig:2014lda}.
  
\subsection{Fermion Singlet DM}
\label{sec:higgsmix}

A simple model including both a real scalar mediator $s$ and a fermion DM singlet $\chi$, which couple through a Higgs portal is given by
\begin{align} \label{eq:LfH}
\mathcal{L}_\text{fermion,$H$} \supset  - \mu_s s^3 - \lambda_s s^4 - y_\chi \bar \chi \chi s - \mu_p s |H|^2 - \lambda_p s^2 |H|^2 \,,
\end{align}
where $y_\chi$ is a Yukawa coupling in the dark sector, while the $\mu_p$ and $\lambda_p$ terms provide the Higgs portal between the dark and the SM sectors. The precise values of the Higgs potential parameters $\mu_s$ and $\lambda_s$ do not play an important role in the DM phenomenology at the LHC and therefore all features relevant for our discussion can be captured within the restricted framework $\mu_s = \lambda_s = 0$. 

In general the Higgs potential in (\ref{eq:LfH}) develops nontrivial VEVs for both $H$ and $s$, but in order to keep the expressions simple it is assumed in the following that $\langle s \rangle = 0$. The main physics implications are unaffected by this assumption. As a result of the portal coupling $\mu_p$, the Higgs and the real scalar fields mix, giving rise to the physical mass eigenstates $h_1$ and $h_2$: 
 \begin{equation}
 \label{eq:seesawInv}
        \begin{pmatrix}
          h_1 \\
          h_2  
        \end{pmatrix}  = 
        \begin{pmatrix}
          \cos \theta &   \sin \theta \\
          - \sin \theta &   \cos \theta    
        \end{pmatrix}
        \begin{pmatrix}
          h \\
         s    
        \end{pmatrix} \, .
      \end{equation}
The mixing  angle is defined such that in the limit $\theta \to 0$ the dark sector is decoupled from the SM. Analytically, one has 
  \begin{align}
      \tan (2 \theta ) = \frac{2v\mu_p}{m_s^2 + \lambda_p v^2 - m_h^2} \,,
  \end{align}
while the masses of $h_1$ and $h_2$ are given by $m_{h_1} \simeq m_h$ and  $m_{h_2} \simeq (m_s^2 + \lambda_p v^2)^{1/2}$. The state $h_1$ can therefore be identified with the bosonic resonance discovered at the LHC. 

To make contact with the  scalar mediator model described in Section~\ref{sec:scalar}, we consider the Yukawa terms that follow from (\ref{eq:LfH}). After electroweak symmetry breaking and rotation to the mass eigenstate basis, one finds 
\begin{equation} \label{eq:LYuk}
{\cal L} \supset  -\frac{1}{\sqrt{2}}\left ( \cos \theta \, h_1 - \sin \theta \, h_2  \right )   \sum_f y_f   \bar f f  -  \left ( \sin \theta \, h_1 + \cos \theta \, h_2  \right)  y_\chi  \bar \chi  \chi \,.
\end{equation}
Identifying $h_2$ with the field $\phi$ in (\ref{eq:scalarlag}), one sees that as far as the couplings between $h_2$ and the SM fermions are concerned, the interactions (\ref{eq:LYuk}) resemble those of the scalar mediator model described in Section~\ref{sec:scalar} with $g_u = g_d = g_e = g_v = -\sin \theta$.  The coupling between DM and the mediator, called $g_\chi$ in (\ref{eq:scalarlag}), is instead given by $g_\chi = y_\chi \cos \theta$. 

Another important feature of  (\ref{eq:LYuk}) is that the effective Yukawa coupling between $h_1$ and the SM fermions is not $y_f$ but  $y_f \cos \theta$. In fact, the  universal suppression factor $\cos \theta$ appears not only in the fermion couplings but also the  $h_1 W^+ W^-$ and $h_1 Z Z$ tree-level vertices as well as the loop-induced $h_1 gg$, $h_1 \gamma \gamma$, and $h_1 \gamma Z$ couplings. The mixing angle and hence (\ref{eq:LfH}) is therefore subject to the constraints that arise from the ATLAS and CMS measurements of the signal strengths in Higgs production and decay. Global fits \cite{Farzinnia:2013pga,Belanger:2013kya} to the LHC Run~I data find $\sin \theta \lesssim 0.4$. Constraints on the mixing angle also derive from the 
oblique
parameters $T$ ({\it aka} the $\rho$ parameter) and~$S$~\cite{Baek:2011aa}, but they are typically weaker  than those that follow from Higgs physics. 

Like in the case of the scalar singlet DM model discussed  before, the model (\ref{eq:LfH}) allows for invisible decays of the Higgs boson, if this is kinematically possible,~i.e.~$m_{h_1} > 2 m_\chi$. The corresponding decay rate is  
\begin{equation} \label{eq:h1invisible}
\Gamma (h_1 \to \chi \bar \chi) = \frac{y_\chi^2 \sin^2 \theta \hspace{0.5mm} m_{h_1}}{8 \pi} \left ( 1 - \frac{4 m_\chi^2}{m_{h_1}^2} \right )^{3/2} \,.
\end{equation}
After the replacements $\sin \theta \to \cos \theta$ and $m_{h_1} \to m_{h_2}$ the same expression holds in the case of $h_2$, if it is sufficiently heavy. In order to determine from (\ref{eq:h1invisible}) the invisible Higgs branching ratio, one has to keep in mind that all partial widths of $h_1$ to SM particles are suppressed by $\cos^2 \theta$ and that depending on the mass spectrum also the decay $h_1 \to h_2  h_2$ may be allowed. 

\begin{figure}[!t]
\begin{center}
\includegraphics[width=0.95\textwidth]{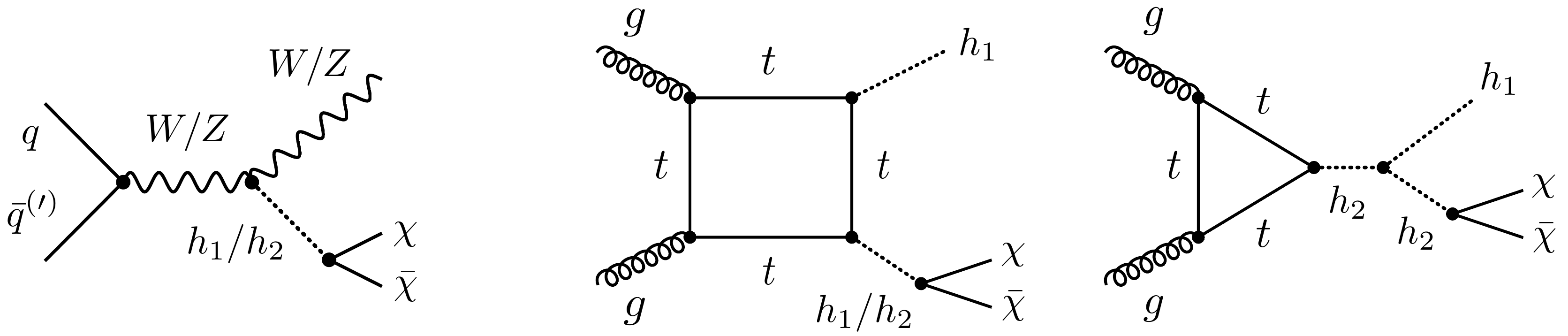} 
\vspace{2mm}
\caption{\label{fig:newsignals} Left: One-loop  
diagrams with an exchange of a $h_1/h_2$ mediator that induces a mono-$W$ and mono-$Z$ signal. Right:  Two possible one-loop graphs that contribute to a mono-Higgs signal. }
\end{center}
\end{figure}

Turning our attention to the $\slashed{E}_T$ signals,  an important observation to make is that the phenomenology of the fermion singlet DM scenario is generically richer than that of the scalar mediator model (\ref{eq:scalarlag}).  First of all, since the Lagrangian (\ref{eq:LfH}) leads to couplings between the scalars $h_1$ and $h_2$ to massive gauge bosons as well as DM pairs, mono-$W$ and mono-$Z$ signals will arise at tree level. The relevant diagrams are shown on the left-hand side in Figure~\ref{fig:newsignals}. The resulting amplitudes for mono-$W$ or mono-$Z$ production at the LHC take the following schematic form 
\beq \label{eq:AppEWZ}
{\cal A} (pp \to \slashed{E}_T + W/Z) \propto y_\chi \sin (2  \theta ) \left ( \frac{1}{s_{\chi \bar \chi} - m_{h_1}^2 + i m_{h_1} \Gamma_{h_1}} - \frac{1}{s_{\chi \bar \chi} - m_{h_2}^2+ i m_{h_2} \Gamma_{h_2}} \right ) \,, 
\eeq
where $s_{\chi \bar \chi}$ denotes the invariant mass of the DM pair and $\Gamma_{h_1}$ and $\Gamma_{h_2}$ are the total decay widths of the scalars. Note that the contributions from virtual $h_1/h_2$ exchange 
have opposite sign 
in~(\ref{eq:AppEWZ}). This implies that the $\slashed{E}_T + W/Z$  signal cross sections can depend sensitively  on $m_{h_2}$ and $m_\chi$ as well as the cuts imposed in the analysis. The destructive interference between the contributions of the two scalar mediators is also at work for mono-jets and it is well-known~\cite{Kim:2008pp,Baek:2011aa,LopezHonorez:2012kv} that it can be phenomenologically relevant in direct detection.  

A second interesting consequence of (\ref{eq:LfH}) is that this Lagrangian gives rise to a mono-Higgs signal \cite{Carpenter:2013xra,Petrov:2013nia}. Two examples of  Feynman graphs  that provide a contribution are given on the right in  Figure~\ref{fig:newsignals}. Notice that while a $\slashed{E}_T + h$ signal can also arise in the simplified $s$-channel scalar mediator scenario discussed in Section~\ref{sec:scalar}, the presence of the two scalar states $h_1$ and $h_2$ and the existence of trilinear Higgs vertices such as $h_1 h_2^2$ are likely to change the mono-Higgs phenomenology of (\ref{eq:LfH}) compared to (\ref{eq:scalarlag}). 

\subsection{Singlet-Doublet DM}

Singlet-doublet DM scenarios are the simplest example of models where the  interactions between DM and the SM arise from mixing of a singlet  with electroweak multiplets. A fermion singlet $\chi$ and a pair of fermion doublets with opposite hypercharge denoted by $\psi_1 = (\psi_1^0,\psi_1^-)^T$ and  $\psi_2= (\psi_2^+,\psi_2^0)^T$ are introduced. Assuming that the new fields are odd under a $Z_{2}$ symmetry under which the SM fields are even, the Lagrangian reads
\begin{align} \label{eq:LFSD}
\mathcal{L}_\text{fermion,SD} & =  i\left(\bar{\chi}\slashed{\partial}\chi + \bar{\psi}_1 \slashed{D}\psi_1+ \bar{\psi}_2 \slashed{D}\psi_2 \right) - \frac{1}{2} m_S \chi^2  -m_{D}\psi_1\psi_2  \nonumber\\
& \phantom{=} -y_1\chi H \psi_1 -y_2  \chi H^\dagger\psi_2 + {\rm h.c.} \,,
\end{align}
where $D_\mu$ denotes the covariant derivative.  The model generalizes the bino-higgsino sector of the MSSM in the decoupling limit. In fact, the Yukawa couplings $y_1$ and $y_2$ are free parameters, whereas in the MSSM they are related to the $U(1)_Y$ gauge coupling.

After electroweak symmetry breaking, singlet and doublets mix. The physical spectrum consists of a pair of charged particles ($\chi^+, \chi^-$) with mass $m_D$
and three neutral eigenstates defined by $(\chi_1, \chi_2, \chi_3)^T = U (\chi, \psi_1^0, \psi_2^0)^T$, where $U$ is the unitary matrix that diagonalizes the mass matrix
\begin{equation}
\mathcal{M}=\left(\begin{array}{ccc}m_{S} & \frac{y_1v}{\sqrt{2}} & \frac{y_2v}{\sqrt{2}} \\[1mm]
\frac{y_1v}{\sqrt{2}} & 0 & m_{D} \\
\frac{y_2v}{\sqrt{2}} &m_{D} & 0 \end{array}\right)\,.
\end{equation}
The DM candidate is the lightest eigenstate $\chi_1$, whose composition in terms of gauge eigenstates is $\chi_1=U_{11}\chi+U_{12}\psi_1^0+U_{13}\psi_2^0$. In the singlet-doublet scenario, DM couples to the Higgs boson $h$ and the SM gauge bosons through its doublet components. The induced interactions can be read off from
\begin{align} \label{eq:LHWZ}
\mathcal{L}\supset& -h\bar{\chi}_i(c^{\ast}_{h\chi_i\chi_j}P_L+c_{h\chi_i\chi_j}P_R)\chi_j-Z_\mu\bar{\chi}_i\gamma^\mu(c_{Z\chi_i\chi_j}P_L-c_{Z\chi_i\chi_j}^{\ast} P_R)\chi_j\nonumber
\\
&-\frac{g}{\sqrt{2}}(U_{i3}W_\mu^-\bar{\chi}_i\gamma^\mu P_L\chi^+ -U_{i2}^{\ast}W_\mu^-\bar{\chi}_i\gamma^\mu P_R\chi^++{\rm h.c.})\,,
\end{align}
where $i,j=1,3$ and
\begin{equation}
c_{Z\chi_i\chi_j}\!=\!\frac{g}{4\cos \theta_w}(U_{i3}U_{j3}^{\ast} -U_{i2}U_{j2}^{\ast} )\,,\qquad
c_{h\chi_i\chi_j}=\frac{1}{\sqrt{2}}(y_1U_{i2} U_{j1}+y_2U_{i3} U_{j1})\,,
\end{equation}
with $g$ the $SU(2)_L$ coupling and $\cos  \theta_w$ the cosine of the weak mixing angle. Due to these interactions, DM can annihilate to SM fermions via $s$-channel Higgs or $Z$-boson exchange and to bosons again through a Higgs or a $Z$ boson in the $s$-channel or via $\chi_i$ or $\chi^+$ in the $t$-channel. Likewise, Higgs ($Z$-boson) exchange leads to SI (SD) DM nucleon scattering. The corresponding phenomenology has been studied in \cite{Enberg:2007rp,Mahbubani:2005pt,Cohen:2011ec,Calibbi:2015nha}. 

As in the case of the other Higgs portal models, a possible collider signature is the invisible width of the Higgs, if the decay $h\to\chi_1\chi_1$
is kinematically allowed:
\begin{equation}
\Gamma(h \to \chi_1 \chi_1)= \frac{m_h}{4 \pi} \left(1 -\frac{4 m_{\chi_1}^2}{m_h^2}\right)^{3/2}
\left| c_{h \chi_{1} \chi_{1} } \right|^2\,.
\end{equation}
Since the $Z$ boson couples directly to pairs of DM particles $\chi_1$, the model (\ref{eq:LFSD}) will also give rise to an additional contribution to the invisible decay width of the $Z$ boson of the form 
\begin{equation}
\Gamma(Z \to \chi_1 \chi_1)= \frac{m_Z}{6 \pi} \left(1 -\frac{4 m_{\chi_1}^2}{m_Z^2}\right)^{3/2}
\left| c_{Z \chi_{1} \chi_{1} } \right|^2\,,
\end{equation}
if $m_Z > 2 m_{\chi_1}$. This possibility  is constrained by the $Z$-pole measurements performed at LEP~\cite{ALEPH:2005ab}, which require 
 $\Gamma(Z \to \chi_1 \chi_1) \lesssim 3 \, {\rm MeV}$.

Since the model (\ref{eq:LFSD})  contains one charged and two neutral fermions in addition to the DM state $\chi_1$, LHC searches for electroweak Drell-Yan production allow to set bounds on the new fermions arising in scalar-doublet scenarios. The relevant production modes are  $q \bar q \to \chi_i \chi_j$ and  $q \bar q \to \chi^+ \chi^-$ via a $Z$ boson or $q \bar q^{(\prime)} \to  \chi^\pm \chi_i$  through $W$-boson exchange. Generically, the latter production mode has the most relevant LHC constraints. Production in gluon-gluon fusion $gg \to \chi_i \chi_i$ through an intermediate Higgs produced via a top-quark  loop is also possible. Like in the case of electroweakino production in the MSSM, final states involving leptons and $\slashed{E}_T$  provide the cleanest way to  probe singlet-doublet models \cite{Enberg:2007rp,Mahbubani:2005pt,Calibbi:2015nha}. A particularly promising channel is for instance $pp \to \chi^\pm \chi_{2,3} \to W^\pm \chi_1 Z \chi_1$ that leads to both a $2 \ell +\slashed{E}_T$ and $3 \ell +\slashed{E}_T$ signature. The scenario (\ref{eq:LFSD}) predicts further collider signals with $\slashed{E}_T$ such as mono-jets that await explorations. 
 
\section{Vector $\bm{s}$-Channel Mediator}
\label{sec:vector}

\subsection{Model-Building Aspects}

One of the simplest ways to add a new mediator to the SM  is by extending its gauge symmetry by a new $U(1)^\prime$, which is spontaneously broken such that the mediator obtains a mass $M_V$~\cite{Holdom:1985ag,Babu:1997st}. Depending on whether DM is a Dirac fermion $\chi$ or a complex scalar $\varphi$, the interactions this new spin-1 mediator take the form~\cite{Dudas:2009uq,Fox:2011qd,Frandsen:2012rk,Alves:2013tqa,Arcadi:2013qia,Lebedev:2014bba}
\begin{align}
{\cal L}_{{\rm fermion}, V} & \supset  V_\mu \hspace{0.25mm} \bar{\chi}  \gamma^\mu (g^V_{\chi}-g^A_{\chi}\gamma_5) \chi +  \sum_{f=q, \ell, \nu }V_\mu \hspace{0.25mm} \bar{f} \gamma^\mu (g^V_{f}-g^A_{f} \gamma_5)f \,, \label{eq:LfV} \\[1mm]
{\cal L}_{{\rm scalar}, V} & \supset  i g_{\varphi}  V_\mu  (\varphi^\ast  \partial^\mu \varphi - \varphi \partial^\mu \varphi^\ast) +  \sum_{f=q, \ell, \nu }V_\mu \hspace{0.25mm} \bar{f} \gamma^\mu (g^V_{f}-g^A_{f} \gamma_5)f \,, \label{eq:LsV}
\end{align}
where $q,\ell$ and $\nu$ denote all quarks, charged leptons and neutrinos, respectively. Under the MFV assumption the couplings of $V$ to the SM fermions will be flavor independent, but they can depend on chirality (such that $g^A_f \neq 0$). For Majorana DM, the vector coupling $g^V_{\chi}$ vanishes, while a real scalar cannot have any CP-conserving interactions with $V$.

In the literature, one often finds a distinction between {\it vector} mediators with vanishing axialvector couplings ($g^A_f = 0$) and {\it axialvector} mediators with vanishing vector couplings ($g^V_f = 0$). Neglecting the couplings to neutrinos, the relevant parameters  in the former case are 
\begin{equation} \label{eq:set1}
\left\{m_\chi, \,M_V,\,g^V_{\chi},\,g^V_u,\,g^V_d,\,g^V_\ell\right\} \; ,
\end{equation}
while, in the latter case, the corresponding set is
\begin{equation} \label{eq:set2}
\left\{m_\chi,\,M_V,\,g^A_{\chi},\,g^A_u,\,g^A_d,\,g^A_\ell\right\} \; .
\end{equation}
Note, however, that it is rather difficult to engineer purely axialvector couplings to all quarks, while being consistent with the SM Yukawa interactions and MFV (as explained below). In the following, we will consider the general case with non-zero vector and axialvector couplings. Although in this case the spin-1 mediator is not a parity eigenstate, we will refer to it as a vector mediator for simplicity.

\subsubsection{The Higgs Sector}

The most straightforward way to generate the mass of the vector mediator is by introducing an additional dark Higgs field $\Phi$ with a non-zero VEV.  Generically, this particle will not couple directly to SM fermions, but it could in principle mix with the SM Higgs, leading to a phenomenology similar to that of Higgs portal models described in Section~\ref{sec:higgsportal}. The mass of the dark Higgs cannot be very much heavier than that of the vector mediator, and so $\Phi$ may need to be included in the description if $M_V$ is small compared to the typical energies of the collider.

Moreover, if the theory is chiral,~i.e.~if $g^A_\chi \neq 0$, the dark Higgs will also be responsible for generating the DM mass. In order for the Yukawa interaction $\Phi\hspace{0.25mm} \bar{\chi}\chi$ to be gauge invariant, we have to require that the $U(1)'$ charge of the left-handed and the right-handed component of the DM field  differ by exactly $q_L - q_R = q_\Phi$. Consequently, the axialvector coupling of DM to the mediator will necessarily be proportional to $q_\Phi$. The longitudinal component of $V$ (i.e.~the would-be Goldstone mode) then couples to $\chi$ with a coupling strength proportional to $g^A_\chi \hspace{0.25mm} m_\chi / M_V$. Requiring this interaction to remain perturbative gives the bound
\begin{equation}
 m_\chi \lesssim \frac{\sqrt{4\pi}}{g^A_\chi} M_V \, ,
 \label{eq:pertYuk}
\end{equation}
implying that the DM mass cannot be raised arbitrarily compared to the mediator mass.

A similar consideration also applies in the visible sector. If the axialvector couplings to the SM states $g^A_{f}$ are non-zero, the only way to have SM Yukawa couplings is if the SM Higgs doublet $H$ carries a charge $q_H$ under the new gauge group. This charge must satisfy $g^\prime  q_H = - g^A_u = g^A_d = g^A_e$ (where $g^\prime$ is the gauge coupling of the $U(1)'$) in order for quark and charged lepton masses to be consistent with the $U(1)'$ symmetry. However, having $q_H \neq 0$ generically implies corrections to electroweak precision measurements, so that one must require $M_V \gtrsim 2$~TeV for consistency with low-energy data.

\subsubsection{Mixing with SM Gauge Bosons}

As soon as there are fermions charged under both the SM gauge group and the new $U(1)'$, loop effects will  induce mixing between the new vector mediator and the neutral SM gauge bosons, in particular kinetic mixing of the form
\begin{equation}
\mathcal{L}_{\rm kinetic} \supset \frac{\epsilon}{2} \, F'^{\mu\nu} B_{\mu \nu} \, ,
\end{equation}
where $F'_{\mu \nu} = \partial_\mu V_\nu - \partial_\nu V_\mu$  and $B_{\mu \nu} = \partial_\mu B_\nu - \partial_\nu B_\mu$ denote the $U(1)'$   and $U(1)_Y$ field strength tensors. Parametrically, this mixing is given by
\begin{equation}
\epsilon \sim \sum_q \frac{(g^A_q)^2}{16 \pi^2} \sim 10^{-2} \, (g^A_q)^2 \, .
\end{equation}
If $M_V$ is too close to the  $Z$-boson mass $M_Z$, this mixing can lead to conflicts with electroweak precision observables~\cite{Carone:1995pu,Babu:1997st,Chun:2010ve,Frandsen:2011cg}. For example, the correction to the $\rho$ parameter, $\Delta \rho = M_W^2 / (M_Z^2  \cos^2 \theta_w) - 1$,  can be estimated to be
\begin{equation}
\Delta\rho \sim \epsilon^2 \frac{M_Z^2}{M_V^2 - M_Z^2} \; .
\end{equation}
Requiring $\Delta\rho \lesssim 10^{-3}$ \cite{Agashe:2014kda} then implies  $g^A_q \lesssim 1$ and $M_V \gtrsim 100\:\text{GeV}$.

\subsection{Phenomenological Aspects}

The first observation is that in models with s-channels mediators, the possibility 
for such particles to decay back to the SM is unavoidably present. This can show up 
as di-jets~\cite{Alves:2013tqa} or di-leptons at the LHC. Indeed the leptonic couplings 
$g^{V}_\ell$ and $g^{A}_\ell$ are very tightly constrained by searches for di-lepton 
resonances~\cite{Arcadi:2013qia, Lebedev:2014bba}.
If the quark couplings of the mediator are equally small, it becomes very difficult to have sizable interactions between the SM and DM and there would typically be no observable DM signals. We therefore focus on the case where the quark couplings of the vector mediator are much larger than the lepton couplings, for example because the SM quarks are charged under the new $U(1)'$ while couplings to leptons only arise at loop-level (a so-called {\it leptophobic} $Z'$ boson).

For such a set-up to be theoretically consistent we must require additional fermions charged under the $U(1)'$ and the SM gauge group to cancel anomalies. The masses of these additional fermions are expected to be roughly of the order of $M_V$, so they can often be neglected in phenomenology, unless the mass of the vector mediator is taken to be small compared to the typical energy scales of the collider. Indeed, it is possible to construct anomaly-free models with no direct couplings to leptons (for example in the context of a baryonic $Z'$ boson~\cite{Duerr:2013lka, Duerr:2014wra}). In this case, the leptonic couplings will not give a relevant contribution to the DM phenomenology of the model and one can simply set $g_\ell^V = g_\ell^A =0$. 

\subsubsection{Collider Searches}

If the vector mediator is kinematically accessible at the LHC, the resulting phenomenology depends crucially on its decay pattern. For  arbitrary vector and axialvector couplings, one finds in the case of Dirac DM the following expression for the total width: 
\begin{align} \label{eq:GammaV}
\Gamma_V =\frac{M_V}{12\pi} \sum_{i=f,\chi} N_c^i \left ( 1-\frac{4m_i^2}{M_V^2} \right )^{1/2}  \, \Big [(g^{V}_{i})^2+(g^{A}_{i})^2 + \frac{m_i^2}{M_V^2} \hspace{0.25mm} \Big (2 \hspace{0.25mm} (g^{V}_{i} )^2 - 4 \hspace{0.25mm} (g^{A}_{i} )^2 \Big)  \Big] \,. 
 \end{align}
Here the sum extends over all fermions $i$ that are above threshold, while $N_c^i = 3$ for quarks and $N_c^i = 1$ for leptons and DM. 

There are several important conclusions that can be drawn from (\ref{eq:GammaV}). The first one concerns the maximal size that the couplings can take to be consistent with $\Gamma_V / M_V < 1$, which is a necessary requirement in order for a perturbative description of the mediator to be valid. Assuming that $M_V \gg m_i$ and setting for simplicity $g_q^V = g_\chi^V = g$ and $g_\ell^V = g_i^A = 0$, one finds that  $\Gamma_V/M_V \simeq 0.5 \hspace{0.25mm} g^2$. This implies that one has to have $g \lesssim 1.4$ in order for the width of the mediator to be smaller than its mass and values of $g$ significantly below unity for the NWA (which calls for $\Gamma_V/M_V \lesssim 0.25$) to be applicable. 

In cases where the NWA can be used, production and decay factorize such that for instance $\sigma (pp \to Z+\chi \bar \chi) = \sigma (pp \to Z+V) \hspace{0.25mm} \times {\rm Br} (V \to \chi \bar \chi)$. The resulting LHC phenomenology is thus determined to first approximation by the leading decay mode of the vector mediator. Considering  a situation with $M_V \gg m_i$ and $g_\ell^V = g_i^A = 0$, one finds that decays into quarks dominate if $g_\chi^V / g_q^V \lesssim 4$, while invisible decays dominate if $g_\chi^V / g_q^V \gtrsim 4$. For $g_\chi^V / g_q^V \simeq 4$ both decay channels have comparable branching ratios. If invisible decays dominate, the strongest collider constraints are expected from searches for $\slashed{E}_T$ in association with SM particles. To illustrate this case, we discuss mono-jet searches below. If, on the other hand, the invisible branching ratio is small, we expect most of the mediators produced at the LHC to decay back into SM particles. In this case, strong constraints can be expected from searches for heavy resonances, and we focus on di-jet resonances.

\subsubsection*{Mono-Jets}

\begin{figure}[!t]
\begin{center}
\includegraphics[width=0.95\textwidth]{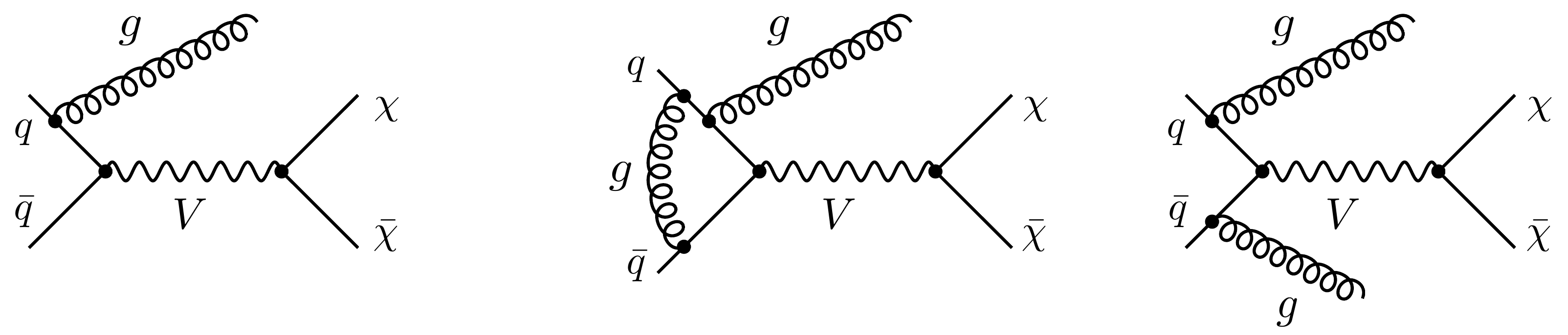} 
\vspace{2mm}
\caption{\label{fig:spin1} Left: An example of a LO diagram  that leads to mono-jet  events through the $s$-channel exchange of a spin-1 vector resonance $V$. Right: At the NLO level both virtual and real corrections have to taken into account in order to obtain a infrared finite result. }
\end{center}
\end{figure}

LHC searches for $\slashed{E}_T$ plus jet signals place strong constraints on the interactions between quarks and DM mediated by a vector mediator~\cite{An:2012va,Frandsen:2012rk,Buchmueller:2013dya,Buchmueller:2014yoa,Harris:2014hga,Fairbairn:2014aqa,Jacques:2015zha,Chala:2015ama}. 
The corresponding cross sections can be calculated at next-to-leading order (NLO) 
in {\tt MCFM}~\cite{Fox:2012ru} and at NLO  plus parton shower 
in the {\tt POWHEG~BOX}~\cite{Haisch:2013ata}. Some of the relevant diagrams are shown in Figure~\ref{fig:spin1}.
If the mediator is too heavy to be produced on-shell at the LHC and assuming equal vector couplings of the mediator to all quarks as well as $g_\ell^V = g_i^A = 0$, the mono-jet cross section at the LHC is proportional to $(g_q^V)^2 \, (g_\chi^V)^2$. The same scaling applies if the mediator is forced to be off-shell because $M_V < 2 m_\chi$ so that decays into DM are kinematically forbidden. 

For $2 m_\chi \ll M_V \ll \sqrt{s}$, with $\sqrt{s}$ the center-of-mass of the collider,
the mediator can be produced on-shell and subsequently decay into a pair of DM particles. If the mediator width is small enough for the NWA to be valid, the mono-jet cross section will be proportional to the product $(g_q^V)^2 \, \text{Br}(V \rightarrow \chi \bar \chi)$. If we fix the ratio $g_\chi^V / g_q^V$, the invisible branching ratio will be independent of an overall rescaling of the couplings, so that we simply obtain $\sigma(pp \rightarrow \slashed{E}_T + j) \propto (g_q^V)^2$. If we rescale only one of the couplings, on the other hand, the resulting change in the mono-jet cross section will depend on the dominant decay channels of the mediator. If the total width of the mediator is dominated by its decays into quarks, the mono-jet cross section will be invariant under a rescaling of the quark coupling $g_q^V$, because the change in the production cross section is compensated by the change in the invisible branching ratio. If, on the other hand, invisible decays dominate, both the production cross section and the invisible branching ratio will be invariant under a (small) change in the coupling $g_\chi^V$.

The same general considerations apply for axialvector couplings instead of vector couplings. In particular, the production cross section of the vector mediator is largely invariant under the exchange $g_q^V \leftrightarrow g_q^A$. Note, however, that for $m_\chi \rightarrow M_V/2$ the phase space suppression is stronger for axialvector couplings than for vector couplings, such that for $m_\chi \simeq M_V/2$ the monojet cross section is somewhat suppressed for a mediator with purely axialvector couplings.

In many situations invisible decays and decays into quarks will both lead to a non-negligible contribution to $\Gamma_V$ as given in (\ref{eq:GammaV}) and furthermore this width may become so large that one cannot use the NWA to derive simple scaling laws. If $m_\chi$ becomes close to $M_V / 2$ there can also be contributions from both on-shell and off-shell mediators. As a result, all relevant parameters ($m_\chi$, $M_V$, $g_\chi^V$ and $g_q^V$) must in general be  taken into account in order to calculate mono-jet cross sections.

\subsubsection*{Di-Jets}

Searches for di-jet resonances exploit the fact that any mediator produced from quarks in the initial state can also decay back into quarks, which lead to observable features in the distribution of the di-jet invariant mass and their angular correlations. However, for small mediator masses the QCD background resulting from processes involving gluons in the initial state completely overwhelms the signal. The most recent di-jet searches at the LHC therefore focus mostly on the region with di-jet invariant mass $m_{jj} \gtrsim 1\:\text{TeV}$. For smaller mediator masses, the strongest bounds are in fact obtained from searches for di-jet resonances at UA2 and the Tevatron~\cite{Chala:2015ama}.  An interesting opportunity to make progress with the LHC even in the low-mass region is to consider the production of di-jet resonances in association with other SM particles, such as $W$ or $Z$ bosons, which suffer from a significantly smaller QCD background~\cite{An:2012ue,Chiang:2015ika}.

An important complication concerning searches for di-jet resonances results from the fact that the width of the mediator can be fairly large. The steeply falling parton distribution functions then imply that the resonance will likely be produced at lower masses, leading to a significant distortion of the expected distribution of invariant masses $m_{jj}$. Existing searches for narrow resonances therefore typically do not apply to vector mediators with couplings of order unity. Nevertheless, the shape of the resonance can still be distinguished from SM backgrounds and it is still possible to constrain these models using specifically designed searches~\cite{Chala:2015ama}.

A number of such searches have been considered in~\cite{Chala:2015ama}. The central conclusion is that, at least for $g_q^V \lesssim 1$, bounds on $M_V$ become stronger as $g_q^V$ is increased, because the enhancement of the production cross section is larger than the reduction of the detection efficiency resulting from the increasing width. Indeed, there are still stringent bounds on mediators with width as large as $\Gamma_V \sim M_V/2$. It is {\it crucial} to take these bounds into account when interpreting DM searches at the LHC in terms of simplified models with an $s$-channel vector mediator, because they apply to a wide range of models and in many cases complement or even surpass other search strategies. A promising strategy to constrain even broader resonances may be to study di-jet angular correlations, such as the ones considered in the context of constraining four-fermion operators (see for instance~\cite{Khachatryan:2014cja, deVries:2014apa}). 

\subsubsection{Direct Detection}

Depending on the coupling structure of the vector mediator, the interactions of DM with nuclei can proceed via SI or SD scattering off nucleons. The corresponding cross sections at zero-momentum transfer are given by
\begin{equation}
\sigma^\text{SI}_{\chi -N} = \frac{\mu^2_{\chi-N}}{\pi M_V^4} \hspace{0.25mm} f_N^2 \,, \qquad 
\sigma^\mathrm{SD}_{\chi -N} = \frac{3 \mu^2_{\chi-N}}{\pi  M_V^4}  \hspace{0.25mm}a_N^2  \,,
\end{equation}
where $N$ stands for either $p$ or $n$, while  $f_N$ and $a_N$ denote the effective nucleon couplings. They take the form 
\begin{equation}
f_p = g^\mathrm{V}_\chi (2 g^V_u + g^V_d) \,, \qquad f_n = g^V_\chi (g^V_u + 2 g^V_d) \,, 
\end{equation}
and
\begin{align}
a_{p,n} = g^A_\chi \sum_{q=u,d,s} \Delta q^{(p, n)} \, g^A_q \, .
\end{align}
The coefficients $\Delta q^{(N)}$ encode the contributions of the light quarks to the nucleon spin. They are given by~\cite{Agashe:2014kda}
\begin{align}
\Delta u^{(p)} & = \Delta d^{(n)} = 0.84 \pm 0.02 \; , \nonumber \\
\Delta d^{(p)} & = \Delta u^{(n)} = -0.43 \pm 0.02 \; , \label{eq:deltas} \\
\Delta s^{(p)} & = \Delta s^{(n)} = -0.09 \pm 0.02 \; . \nonumber
\end{align}
Potential cross terms such as $g^V_q g^A_\chi$ are suppressed in the non-relativistic limit (either by the momentum transfer or the DM velocity, both of which lead to a suppression of $10^{-3}$ or more), and can therefore be neglected.

Substituting the expressions for the effective couplings into the formulas for the DM-nucleon scattering cross sections, we obtain
\begin{eqnarray}
\sigma^\mathrm{SI}_{\chi-N} & \! = \! & 1.4\times10^{-37}\text{ cm}^2\,g^V_\chi g^V_q\,{\left(\frac{\mu_{\chi-N}}{1 \, \text{GeV}}\right)}^2{\left(\frac{300 \, \text{GeV}}{M_V}\right)}^4, \\[.5cm]
\sigma^\mathrm{SD}_{\chi-N} & \! = \! & 4.7\times10^{-39}\text{ cm}^2\,g^A_\chi g^A_q\,{\left(\frac{\mu_{\chi-N}}{1 \, \text{GeV}}\right)}^2{\left(\frac{300 \, \text{GeV}}{M_V}\right)}^4 \; .
\end{eqnarray}
Crucially, SI interactions receive a coherent enhancement proportional to the square of the target nucleus mass, leading to very strong constraints from direct detection experiments unless the DM mass is very small. Consequently, the estimates above imply that for $g_q \simeq 1$, SI interactions are sensitive to mediator masses of up to $M_V \simeq 30\:\text{TeV}$, while SD interactions only probe mediator masses up to around $M_V \simeq 700\:\text{GeV}$. This should be contrasted with the constraints arising from the LHC, which are close to identical for vector and axialvector mediators. 

\subsubsection{Annihilation}

Two processes contribute to DM annihilation in the early Universe: annihilation of DM into SM fermions and (provided $M_V \lesssim m_\chi$) direct annihilation into pairs of mediators, which subsequently decay into SM states. For the first process, the annihilation cross section is given by
\begin{align}
(\sigma v) (\chi \bar{\chi} \to V \rightarrow q \bar{q}) &=  \frac{3 m_\chi^2}{2 \pi \left[(M_V^2 - 4 m_\chi^2)^2 + \Gamma_V^2 M_V^2\right]}  \left ( 1 - \frac{4m_q^2}{M_V^2} \right )^{1/2} \nonumber \\[1mm] & \hspace{-3cm} \times \left \{ 
 (g^V_\chi)^2 \left[(g^V_q)^2 \left (2 + \frac{m_q^2}{M_V^2} \right ) + 
2  \hspace{0.25mm} (g^A_q)^2 \left (1 - \frac{m_q^2}{M_V^2} \right )  \right] +
(g^A_q)^2 (g^A_\chi)^2 \frac{m_q^2}{M_V^2} \frac{(4 m_\chi^2 - M_V^2)^2}{M_V^4} \right \}  \, ,
\label{eq:ann}
\end{align}
where $\Gamma_V$ is the total decay width of the vector mediator as given in (\ref{eq:GammaV}). For $m_\chi \simeq M_V / 2$ the annihilation rate receives a resonant enhancement, leading to a very efficient depletion of DM.

An important observation is that for $g^V_\chi = 0$, the annihilation cross section is helicity-suppressed. For $m_b \ll m_\chi < m_t$ the factor $m_q^2 / m_\chi^2$ can be very small, such that it is important to also include the $p$-wave contribution for calculating the DM relic abundance. Including terms up to second order in the DM velocity $v$, we obtain for the special case  $g^V_q = g^V_\chi = 0$ the expression 
\begin{align}
(\sigma v) (\chi \bar{\chi} \to V \rightarrow q \bar{q})  & =  \frac{(g^A_q)^2 (g^A_\chi)^2  \hspace{0.5mm} m_\chi^2}{2 \pi \left[(M_V^2 - 4 m_\chi^2)^2 + \Gamma_V^2 M_V^2\right]}  \left ( 1 - \frac{4m_q^2}{M_V^2} \right )^{1/2} \nonumber \\[1mm] & \phantom{=} \times \left \{ 
\frac{3 m_q^2}{M_V^2} \frac{(4 m_\chi^2 - M_V^2)^2}{M_V^4} + \left ( 1 - \frac{m_q^2}{M_V^2} \right ) v^2 \right \} \,.
\end{align}

Finally, the annihilation cross section for direct annihilation into pairs of mediators is given by
\begin{align}
 (\sigma v)(\chi \bar{\chi} \rightarrow V V) & =  \frac{(m_\chi^2 - M_V^2)^{3/2}}{4 \pi  \hspace{0.25mm} m_\chi M_V^2 (M_V^2 - 2 m_\chi^2)^2} \nonumber \\[1mm] & \phantom{=} \times \Bigl( 8 (g^A_\chi)^2 (g^V_\chi)^2 m_\chi^2 + \left[ (g^A_\chi)^4 - 6 (g^A_\chi)^2 (g^V_\chi)^2 + 
      (g^V_\chi)^4 \right] M_V^2 \Bigr) \, .
\end{align}

We note that for the coupling strengths and mass ranges typically considered in the context of LHC DM searches, it is easily possible to achieve sufficiently large annihilation cross sections to deplete the DM abundance in the early Universe. In fact, the generic prediction in large regions of parameter space would be that the DM particle is underproduced. In this case, the observed DM relic abundance can still be reproduced if one assumes an initial particle-antiparticle asymmetry in the dark sector, such that only the symmetric component annihilates away and the final DM abundance is set by the initial asymmetry.

\section{$\bm t$-Channel Flavored Mediator}
 \label{sec:tchannel}
 
 If the DM is a fermion $\chi$, the mediator can be a colored scalar or a vector particle $\phi$.  We focus on the scalar case, which makes contact with the MSSM
 and is easier to embed into a UV-complete theory.  A coupling of the form $\phi\bar{\chi} q$ requires either $\chi$ or $\phi$ to carry a flavor index in
 order to be consistent with MFV.  We choose the case where the colored scalar $\phi$ carries the flavor index (much like in the MSSM case, where the colored
 scalar quarks come in the same flavors as the SM quarks).  This class of models has been considered previously in~\cite{Chang:2013oia,Bai:2013iqa,Bell:2012rg,An:2013xka,DiFranzo:2013vra,Papucci:2014iwa,Busoni:2014haa}, while models where  $\chi$ carries the flavor index have been studied in~\cite{Agrawal:2011ze,Kile:2013ola,Agrawal:2014una}.
 
 There are variations where the mediator couples to right-handed up-type quarks, right-handed down-type quarks, or left-handed quark doublets.  For definiteness,
 we discuss the right-handed up-type case (the other cases are obtained in a similar fashion).  In this case, there are three mediators
 $\phi_i = \left\{ \tilde{u}, \tilde{c}, \tilde{t} \right\}$, which couple to the SM and DM via the interaction
 \bea \label{eq:Ltchannel}
{\cal L}_{{\rm fermion}, \tilde u} \supset  \sum_{i=1,2,3} g \hspace{0.25mm} \phi_i^* \bar{\chi} P_R u_i + {\rm h.c.} 
 \eea
 Note that MFV requires both the masses $M_{1,2,3}$ of the three mediators to be equal and universal couplings $g=g_{1,2,3}$ between the mediators and their corresponding quarks $u_i = \left\{ u, c, t \right\}$. This universality can however be broken by allowing for corrections to (\ref{eq:Ltchannel}) and the mediator masses which involve a single insertion of the flavor spurion $Y^{u\, \dagger} Y^u$. Because of the
 large top-quark Yukawa coupling,  in this way  the mass of the third mediator and its coupling can be split 
 from the other two.  In practice this means that the generic parameter space is five-dimensional: 
 \bea
 \left\{ m_\chi,~ M_{1,2},~ M_3,~ g_{1,2},~ g_3 \right\} \,.
 \eea
 These simplified models are very similar to the existing ones for squark searches \cite{Alves:2011wf}, and results can often be translated from one to the other with relatively
 little work.  Note that most studies will involve $g_{1,2}$ together with $M_{1,2}$ {\em or} $g_3$ together with $M_3$.  So specific applications will often have a smaller
 dimensional space of relevant parameters. In the discussion below, we restrict attention to the parameter space with $g_{1,2}$, $M_{1,2}$, and $m_\chi$.  For models where $g_3$ and $M_3$ are relevant, see~\cite{Kumar:2013hfa,Batell:2013zwa,Agrawal:2014una,Kilic:2015vka}.

\subsection{Collider Constraints}
 
Given the masses and couplings, the widths of the mediators are calculable. One finds 
\begin{align}
\Gamma(\phi_i \to \chi \bar u_i) &=
\frac{g_i^2}{16\pi M_i^3}(M_i^2-m_{u_i}^2-m_{\chi}^2) \nonumber \\& \phantom{=} \times \sqrt{M_i^4 + m_{u_i}^4 + m_{\chi}^4 - 2M_i^2 m_{u_i}^2 - 2M_i^2 m_{\chi}^2 - 2m_{\chi}^2 m_{u_i}^2} \nonumber \\[2mm]
& = \begin{cases} \frac{g_i^2}{16\pi} M_i \left (1-\frac{m_{\chi}^2}{M_i^2} \right )^2 \,, & \; M_i, m_{\chi} \gg m_{u_i} \,. \\ 
\frac{g_i^2}{16\pi} M_i  \,, & \; M_i \gg m_{\chi}, m_{u_i} \,. \end{cases} 
\end{align}
Unless the final-state quark $u_i$ is a top quark, the given limiting cases are always very good approximations to the exact widths. 

\begin{figure}[!t]
\begin{center}
\includegraphics[width=0.95\textwidth]{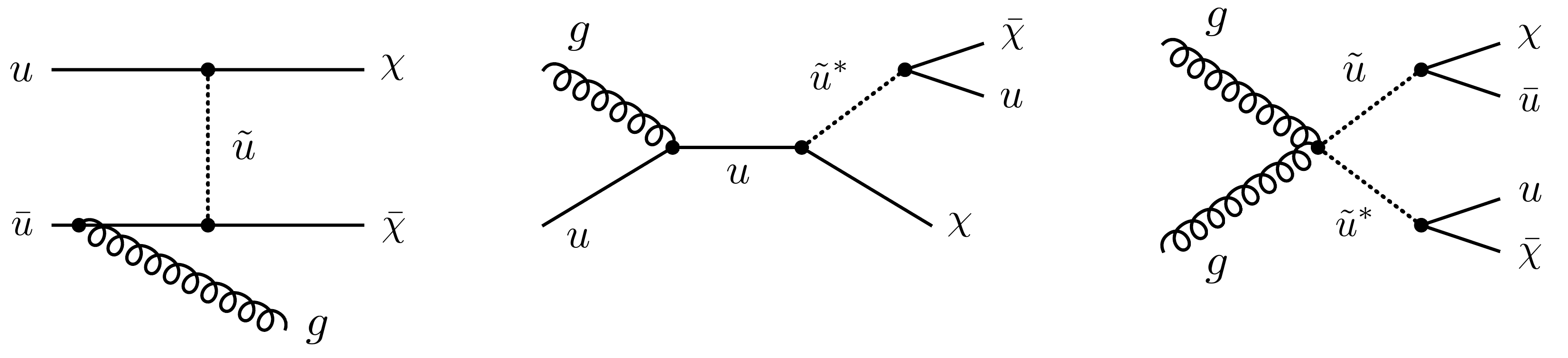} 
\vspace{2mm}
\caption{\label{fig:tchannel} A $\slashed{E}_T + j$ signal can arise in the $t$-channel mediator scenario from initial-state gluon emission (left) and associated mediator production  (middle). Initial-state gluon splitting processes and gluon emission from the $t$-channel mediator is also possible but  not  shown. Pair production of the mediator $\tilde u$ in gluon fusion leads instead to $\slashed{E}_T + 2j$ events (right). Quark-fusion pair production either via  $s$-channel gluon or $t$-channel DM exchange also contributes to the latter signal. 
}
\end{center}
\end{figure}

In the context of (\ref{eq:Ltchannel}) the  production channels that lead to a $\slashed{E}_T + j$ signal are $u \bar u \to \chi \bar \chi + g$, $ug  \to \chi \bar \chi + u$ and $\bar ug  \to \chi \bar \chi + \bar u$. Examples of the relevant Feynman diagrams are shown on the left and in the middle of Figure~\ref{fig:tchannel}. In addition, if the colored mediator $\tilde u$ is sufficiently light it may be pair produced  from both $gg$ or $u \bar u$ initial states. This gives rise to a $\slashed{E}_T + 2j$ signature as  illustrated by the graph on the right-hand side in the same figure.  If the DM particle is a Majorana fermion also the $uu$ and $\bar u \bar u$ initial states contribute to the production of mediator pairs. The latter corrections vanish if $\chi$ is a Dirac fermion. From this brief discussion, it should be clear that $t$-channel models can be effectively probed through both 
mono-jet and squark searches.

\subsubsection{Mono-Jet Searches}

Given that in all recent mono-jet analyses a second hard jet is allowed, the corresponding LHC searches are sensitive in $t$-channel models to the contributions not only from initial-state gluon radiation and associated production, but also to mediator pair production. Since the relative importance of the different channels depends on $m_\chi$, $M_1$, and $g_1$ as well as the imposed experimental cuts,  all corrections should be included in an actual analysis. General statements about the leading partonic channel are however possible. For what concerns $\slashed{E}_T + j$ events the diagram in the middle of Figure~\ref{fig:tchannel} usually gives the dominant contribution.  Compared to $u \bar u \to \chi \bar \chi + g$, this process benefits from a phase-space  enhancement, the larger gluon luminosity, and  the fact that  jets from initial state radiation tend to be softer.  If the mass $M_1$ is small, diagrams with gluon emission from the mediator can also be important, but these graphs are subdominant if the mediator is heavy, since they are $1/M_1^{2}$ suppressed. Notice that the dominance of the associated production channel is a distinct feature of $t$-channel models that is not present in the case of $s$-channel mediators, nor is it relevant in supersymmetric theories where the mediator is a squark. The relative importances of the different $\slashed{E}_T + j$ and $\slashed{E}_T + 2j$  channels depend sensitively on how $g_1$ compares to the strong coupling constant $g_s$. In the limit  $g_1\ll g_s$, pure QCD pair production dominates, while  in the opposite case graphs with DM exchange are more important. Detailed studies of the bounds on the coupling $g_1$ as a function of $M_1$ and $m_\chi$ that arise from  Run~I mono-jet data have been presented in~\cite{An:2013xka,Papucci:2014iwa}. 

\subsubsection{Squark Searches}

If the $t$-channel mediator is light it  can be copiously produced in pairs at the LHC and then decay into DM  and a quark. The resulting phenomenology is very similar to squark pair production in the MSSM with a decoupled gluino.  There is however one important difference which has to do with the fact that in supersymmetric theories the coupling between the squarks and the neutralino~$\chi$ is necessarily weak. The cross section for squark  pair production  through $t$-channel exchange of DM is therefore negligible. This is not the case in $t$-channel mediator scenarios, because $g_1$ is a free parameter and thus it is possible to enhance significantly the  $\tilde u$ pair production rate associated to $t$-channel DM exchange. As already mentioned, the quark-fusion pair production cross section depends on whether $\chi$ is a Dirac or a Majorana particle. In the former case only $u \bar u$-initiated production is non-zero, while in the latter case also the $u u$ and $\bar u \bar u$ initial states furnish a contribution. The constraints from LHC squark searches on $t$-channel mediator models with both Dirac and Majorana DM have been investigated thoroughly in \cite{Bai:2013iqa,An:2013xka,DiFranzo:2013vra,Papucci:2014iwa}. These studies show that squark and mono-jet searches provide comparable and complementary bounds on a wide range of the parameter space of $t$-channel scenarios, depending on the masses of the mediator and DM. Especially in the case where the DM particle and the mediator are quasi-degenerate in mass, mono-jet searches turn out to be superior. 

 \subsection{Scattering with Nucleons}
 
Away from resonance and neglecting light quark-mass effects, the SI scattering cross section of Dirac DM and nucleons that is induced by (\ref{eq:Ltchannel}) reads
 \beq
\sigma_{\chi - N}^{\text{SI}} = \frac{g_1^4 \hspace{0.25mm} \mu_{\chi-N}^2 }{64\pi \left ( M_1^2 - m_{\chi}^2 \right )^2} \,  f_N^2\,.
\eeq
Here $f_p = 2$ and $f_n = 1$ and hence the SI cross sections for protons and neutrons are different in the $t$-channel scenario. Using the same approximations the subleading SD scattering cross section takes the form 
\beq \label{eq:tchSD}
\sigma_{\chi - N}^{\rm SD} = \frac{3 g_1^4 \hspace{0.25mm} \mu_{\chi-N}^2 }{64\pi \left (M_1^2 - m_{\chi}^2 \right )^2} \, \big  ( \Delta u^{(N)} \big )^2 \,,
\eeq
with the numerical values for $\Delta u^{(N)}$ given in (\ref{eq:deltas}). Notice that for Majorana DM, the SI scattering cross section vanishes and the expression for $\sigma_{\chi - N}^{\rm SD}$ is simply obtained from  (\ref{eq:tchSD}) by multiplying the above result  by a factor of 4. 

Since in $t$-channel models with Dirac DM  one has $\sigma_{\chi - N}^{\text{SI}} \neq 0$, the existing direct detection constraints dominate over the collider bounds up to very low DM masses of around $5 \, {\rm GeV}$.  For Majorana DM instead~---~as a result  of the lack of the enhancement from coherence in DM-nucleus scattering~---~the LHC constraints turn out to be stronger than the direct detection limits for  DM masses up to of a few hundred GeV.

 \subsection{Annihilation Rates}

The main annihilation channel of DM in the framework of (\ref{eq:Ltchannel})  is $\chi \bar \chi \to u_i \bar u_i$. In the Dirac case this leads to a $s$-wave contribution of the form 
\beq
(\sigma v) (\chi \bar \chi \to u_i \bar u_i) =\frac{3 g_i^4 \hspace{0.25mm}m_\chi^2 }{32\pi \left (M_i^2 + m_{\chi}^2 \right )^2}  \,,
\eeq
if quark masses are neglected (remember that an additional factor of 1/2  has to be included in the thermal averaging). In the Majorana case, annihilation to SM quarks is instead $p$-wave suppressed and given for zero quark masses by 
\beq
(\sigma v) (\chi \chi \to u_i \bar u_i) = \frac{g_i^4 \hspace{0.25mm} m_\chi^2  (M_i^4 + m_\chi^4)}{16\pi \left (M_i^2 + m_{\chi}^2 \right )^4} \, v^2  \,.
\eeq

In the parameter space where the mediator $\phi_i$ and the DM particle $\chi$ are quasi-degenerate in mass and the ratio $(M_i - m_\chi)/m_\chi$ is comparable to or below the freeze-out temperature,  co-annihilation effects become important~\cite{Griest:1990kh,Edsjo:1997bg}. For both Dirac and Majorana fermions the annihilation cross section for $\chi \phi_i \to u_i g$  can be written as 
\beq
(\sigma v) ( \chi \phi_i^\ast \to  u_i g)  = \frac{g_s^2 g_i^2}{24 \pi M_i \left ( M_i + m_\chi \right ) } \,,
\eeq
if quark-mass effects and $v^2$-suppressed terms are neglected. In addition the mediators $\phi_i$ can self-annihilate. While  for both Dirac and Majorana DM annihilation to gluons 
\beq
(\sigma v) ( \phi_i \phi_i^\ast \to  gg)  = \frac{7 g_s^4}{216 \pi M_i^2 } \,,
\eeq
proceeds via $s$-wave, the process $\phi_i \phi_i^\ast \to  u_i \bar u_i$ is $p$-wave suppressed and hence subdominant. Finally, for Majorana DM the reaction $\phi_i \phi_i \to u_i u_i$ (and its charge conjugate) is possible. The relevant $s$-wave contribution in this case reads 
\beq
(\sigma v) ( \phi_i \phi_i \to  u_i u_i)  = \frac{g_i^4 \hspace{0.25mm} m_\chi^2}{6 \pi \left ( M_i^2 + m_\chi^2 \right )^2} \,.
\eeq

Assuming that the relic abundance $\Omega_\chi h^2$ is thermally produced, one finds that for Dirac DM there is no region in the parameter space that satisfies the combined constraints arising from the LHC searches, direct detection, and $\Omega_\chi h^2$. Therefore the simple model (\ref{eq:Ltchannel}) with Dirac DM cannot be regarded as a complete model in describing the  interactions between the dark and the visible  sectors. In the case of Majorana fermions satisfying all three requirements is possible, but the mass of DM must be larger than about 100 GeV. If DM is lighter there must be other channels for DM to annihilate into, which calls for additional new physics.

\section{Conclusions}
\label{sec:conclusions}

The primary goal of this document is to outline a set of simplified models 
of DM and their interactions with the SM.  It can thus serve as a summary and 
proposal for the simplified models to be implemented in future searches for 
DM at the LHC. The list of models discussed includes spin-0 
and spin-1 $s$-channel  mediator scenarios as well as $t$-channel models. 
The most important prototypes of Higgs-portal scenarios are also described.  
To motivate our choice of  simplified models, a number of 
guiding principles have been given that theories of DM-SM interactions should satisfy in 
order to be useful at LHC energies (and possibly beyond). Based on these 
criteria building further simplified  (or even complete) DM models is 
possible. While the focus is on giving a brief account of the LHC signals 
that seem most relevant in each of the simplified models, we have also 
provided expressions and formulas for reference that allow the reader to 
derive the constraints from direct and indirect searches for DM. There is 
still useful work to be done to improve our understanding of simplified DM 
models, and room to devise creative new searches that can discover or 
constrain them.

While most of the discussion in this work centers around simplified models, 
we emphasize that the given examples represent only ``theoretical sketches''
of DM-SM interactions, and that they by no means exhaust the whole spectrum 
of possibilities that the DM theory space has to offer. They are neither 
meant to form self-contained, complete pictures of DM interactions at the 
LHC, nor are they meant to be model-independent and general enough to cover 
the {\em entirety} of the DM landscape. In order to do justice to the range of 
options in the DM theory space, it is thus important when searching for DM 
at the LHC to frame the results of searches in terms of all three types of theoretical 
frameworks: EFTs, simplified models, and UV complete theories. Only in this 
way is it possible to maximize the search coverage for DM at LHC Run~II, 
and have the largest possible impact on our understanding of the particle 
properties of DM. Simplified models thus play a crucial role in this endeavor.

\section*{Acknowledgments}
The DM@LHC2014 Workshop organizers 
wish to acknowledge the generous support of the UK's Science and Technology 
Facilities Council, the Rutherford Appleton Laboratory, the University of Oxford, 
Imperial College London, IPPP Durham, the Universit\'e de Gen\`eve, and Dark 
Matter Coffee of Chicago.  The final session of the workshop, a discussion 
session highlighting simplified models, was supported by the Institute of 
Physics as a ``Half-Day Meeting''.  
%
In addition we gratefully acknowledge support for staff provided by CERN and the following funding agencies:
ARC (Australia);
FNRS and FWO (Belgium); 
NSERC, NRC, and CFI (Canada);
CAS, MoST and NSFC (China);
CEA and CNRS/IN2P3 (France);
BMBF, DFG, HGF and MPG (Germany); 
INFN (Italy);
ISF (Israel); 
MEXT and JSPS (Japan);
MOE and UM (Malaysia); 
FOM and NWO (The Netherlands);
SNSF and SER (Switzerland);
MST (Taiwan); 
STFC (United Kingdom);
DOE and NSF (USA).




\end{document}